\documentclass{article}
\usepackage{jmlr2e}
\usepackage{amsmath}
\usepackage{bm}
\usepackage{subfig}
\usepackage{changepage}
\usepackage{siunitx}
\DeclareMathOperator{\E}{\mbox{E}}

\DeclareMathOperator{\tr}{tr}
\DeclareMathOperator{\KL}{KL}
\DeclareMathOperator{\diag}{diag}
\renewcommand\vec{\bm}

\ShortHeadings{Nonnegative spatial factorization}{Townes and Engelhardt}
\firstpageno{1}

\begin{document}

\title{Nonnegative spatial factorization}
\author{\name F. William Townes\textsuperscript{1} \and \name Barbara E. Engelhardt\textsuperscript{1,2} \\
\email \{{ftownes,bee\}}@princeton.edu\\
\\
\addr \textsuperscript{1}Department of Computer Science\\
Princeton University\\
Princeton, NJ\\
\\
\textsuperscript{2}Data Science and Biotechnology Institute\\
Gladstone Institutes\\
San Francisco, CA
}

\editor{}

\maketitle

\begin{abstract}
Gaussian processes are widely used for the analysis of spatial data due to their nonparametric flexibility and ability to quantify uncertainty, and recently developed scalable approximations have facilitated application to massive datasets. For multivariate outcomes, linear models of coregionalization combine dimension reduction with spatial correlation. However, their real-valued latent factors and loadings are difficult to interpret because, unlike nonnegative models, they do not recover a parts-based representation. We present nonnegative spatial factorization (NSF), a spatially-aware probabilistic dimension reduction model that naturally encourages sparsity. We compare NSF to real-valued spatial factorizations such as MEFISTO \citep{velten_identifying_2020} and nonspatial dimension reduction methods using simulations and high-dimensional spatial transcriptomics data. NSF identifies generalizable spatial patterns of gene expression. Since not all patterns of gene expression are spatial, we also propose a hybrid extension of NSF that combines spatial and nonspatial components, enabling quantification of spatial importance for both observations and features. A TensorFlow implementation of NSF is available from \url{https://github.com/willtownes/nsf-paper}.
\end{abstract}


\section{Introduction}

Spatially-resolved transcriptomics (ST) has revolutionized the study of intact biological tissues \citep{moses_museum_2021,editors_method_2021,maynard_transcriptome-scale_2021}. In contrast to single-cell RNA sequencing (scRNA-seq), which dissociates cells before sequencing each one, ST quantifies gene expression while preserving the spatial context of the cells within the tissue sample. Since the state and function of each cell is highly dependent upon interactions with its neighbors \citep{verma_self-exciting_2021}, measuring spatially-resolved transcription represents a crucial advance in our ability to understand cellular state and interactions. 

Like scRNA-seq, ST data generally consist of discrete counts of transcripts from tens to thousands of genes, many of which are zero. Also, in both techniques there is typically no ground truth assignment of cell types.
There are two basic strategies to measure spatial gene expression: microscopy and bead capture. Microscopy approaches have excellent spatial resolution, even at the sub-cellular level, but require specialized equipment and may not capture large numbers of cells or genes easily. Examples of microscopy protocols include seqFISH+ \citep{eng_transcriptome-scale_2019} and MERFISH \citep{xia_spatial_2019}. Protocols based on bead capture and sequencing tend to have coarser spatial resolution but cover larger spatial areas or larger numbers of genes. They are popular because they use equipment and experimental procedures that are similar to single-cell RNA-seq (scRNA-seq). Examples of bead protocols include high definition spatial transcriptomics (HDST) \citep{vickovic_high-definition_2019}, Slide-seqV2 \citep{stickels_highly_2021}, and the Visium platform from 10x genomics.

Dimension reduction (DR) is a vital tool for unsupervised learning, and there has been a proliferation of DR methods for both scRNA-seq \citep{wolf_scanpy_2017,butler_integrating_2018,sun_accuracy_2019} and ST \citep{palla_squidpy_2021,dries_giotto_2021}. DR based on a Gaussian error assumption, such as principal components analysis (PCA) \citep{hotelling_analysis_1933} and factor analysis \citep{bartholomew_latent_2011}, is often computationally fast, but requires elaborate normalization procedures that may systematically distort the count data from sequencing technologies \citep{hicks_missing_2018,townes_feature_2019}. For example, this has led to confusion about whether zero-inflated distributions are needed to analyze scRNA-seq data, or whether the high number of zeros is consistent with a simpler Poisson or negative binomial count distribution \citep{svensson_droplet_2020,kim_demystifying_2020,sarkar_separating_2020}. 
To avoid normalization and its pitfalls, DR approaches such as scVI \citep{lopez_deep_2018}, CPLVM~\citep{jones_contrastive_2021}, and GLM-PCA \citep{townes_glmpca:_2019} operate directly on raw counts of unique molecular identifiers (UMIs) by assuming appropriate likelihoods such as the Poisson or negative binomial.

Since ST data, like scRNA-seq, are high-dimensional molecule counts that map onto specific gene transcripts, in principle existing DR methods can be used to obtain a low-dimensional representation of gene expression at each spatial location. This is in fact the standard procedure recommended by tutorials from two popular packages: Seurat\footnote{\url{https://satijalab.org/seurat/articles/spatial_vignette.html}} and Scanpy\footnote{\url{https://scanpy-tutorials.readthedocs.io/en/latest/spatial/basic-analysis.html}}.
However, this application of scRNA-seq methods ignores the spatial coordinates that are the distinguishing feature of ST. We would instead like to retain spatial locality information while performing DR on these data.

Gaussian processes (GPs) are probability distributions over arbitrary functions on a continuous (e.g., spatial) domain \citep{rasmussen_gaussian_2005}. GPs are a fundamental tool in spatial statistics \citep{banerjee_hierarchical_2014,cressie_spatial_2021}. Historically, GPs have been widely used in environmental applications with spatial structure~\citep{finley_efficient_2019}. In the genomics (ST) context, spatialDE \citep{svensson_spatialde_2018} fits univariate GP models to ST data to identify which genes are spatially variable. Other examples of univariate GPs applied to ST data include the Bayesian hierarchical model Splotch \citep{aijo_splotch_2019} and the scalable GPcounts \citep{bintayyash_non-parametric_2021}. While these methods make positive steps toward including spatial information in routine ST analyses, they do not provide dimension reduction. Genes do not act in isolation but interact with each other. This means there is substantial gene-gene correlation in multivariate ST data ignored by univariate approaches. 

A multivariate approach to spatially-aware dimension reduction for ST data is provided by MEFISTO \citep{velten_identifying_2020}. The key concept of MEFISTO is to represent the high-dimensional gene expression features as a linear combination of a small number of independent GPs over the spatial domain. This is known in the statistics literature as a linear model of coregionalization (LMC) \citep{gelfand_nonstationary_2004}. Historically, LMC factorization required a conjugate (Gaussian) likelihood for computational tractability, which is not appropriate for ST count data. Following \cite{yu_gaussian-process_2009}, we refer to this model as Gaussian process factor analysis (GPFA). In neuroscience, attempts were made to relax the Gaussian assumption for application to functional MRI data, leading to count-GPFA \cite{zhao_variational_2017}. 

Even with conjugate likelihoods and univariate outcomes, exact inference for GPs scales cubically with the number of observations (or spatial locations), which is often prohibitive for ST. For example, the recent Slide-seqV2 protocol can generate tens of thousands of observations \citep{stickels_highly_2021}. Breakthroughs in variational inference for GPs have greatly improved scalability and enabled nonconjugate likelihoods through approximate inference with inducing points (IPs; see \cite{leibfried_tutorial_2021} and \cite{van_der_wilk_framework_2020} for overviews). GPFlow is a popular implementation supporting a variety of likelihoods \citep{matthews_gpflow_2017}. MEFISTO also uses the variational IP strategy, and in principle is compatible with nonconjugate likelihoods, but in practice the authors recommend using a Gaussian likelihood \citep{velten_identifying_2020}. An alternative to IPs using polynomial approximate likelihoods \citep{huggins_pass-glm_2017} has been proposed for count-GPFA \citep{keeley_efficient_2020}.

The latent, low-dimensional spatial factors discovered by LMC variants such as GPFA, count-GPFA, and MEFISTO are real-valued. Thus, they may be thought of as spatial analogs of PCA (when the data likelihood is Gaussian) and GLM-PCA (for non-Gaussian data likelihoods). In all cases, latent factors are combined linearly to predict the outcomes. We refer to the weights in these linear combinations as \emph{loadings}, and note that, in the models described above, these loadings are assumed to be real-valued as well. We will use the terms real-valued spatial factorization (RSF) and factor analysis (FA) to refer to these spatial and nonspatial models, respectively. Both RSF and FA models tend to produce dense loadings, but MEFISTO counteracts this with sparsity-promoting priors on the loadings matrix. Sparse loadings are more interpretable than dense because, through nonzero values in the loadings, they assign a small number of relevant features to each component, rather than matching every feature to every component.

Another way to generate sparse loadings is to constrain the entire model to be nonnegative. In the non-spatial context, nonnegative matrix factorization (NMF) \citep{lee_learning_1999} and latent Dirichlet allocation (LDA) \citep{blei_latent_2003} are widely used to produce interpretable low-dimensional factorizations of high-dimensional count data \citep{carbonetto_non-negative_2021} including scRNA-seq \citep{elyanow_netnmf-sc_2020,sherman_cogaps_2020} and ST \citep{zeira_alignment_2021}. To quantify uncertainty, a Bayesian prior can be placed on latent factors and a Poisson or negative binomial data likelihood included to lead to probabilistic NMF (PNMF). The advantage of PNMF over real-valued alternatives is that, for geometric reasons, they produce parts-based representations rather than holistic representations. For example, when decomposing pixel-based representations of faces, PNMF decomposes a face into factors representing eyes, noses, mouths, and ears distinctly~\citep{lee_learning_1999}. On the other hand, real-valued alternatives produce \emph{eigenfaces}, or representations of whole faces in each factor \citep{lee_learning_1999}. 
Incorporating nonnegativity constraints into spatial models is not straightforward, though, since GPs are inherently real-valued.

The contributions of this work are threefold. First, we develop nonnegative spatial factorization (NSF), a model that allows spatially-aware dimension reduction using a Gaussian process prior over the spatial locations and with a Poisson or negative binomial likelihood for count data. Second, we combine this spatially-aware dimension reduction with nonspatial factors in a NSF hybrid model (NSFH) to partition variability into the spatial and nonspatial sources. Finally, we identify appropriate GP kernels and develop inference methods for the kernel parameters and latent variables to enable computationally tractable fitting of large field-of-view ST data.

This paper proceeds as follows. We first define the generative models of FA, PNMF, RSF, NSF, and NSFH. Secondly, we illustrate the ability of nonnegative factorizations to identify a parts-based representation using simulations. We then describe the basic features of the ST datasets and examine key results of a benchmarking comparison of different models. Next, we analyze three ST datasets from different technologies with NSFH and show how to interpret the spatial and nonspatial components. We conclude with a discussion of the implications of our results in ST data analysis and promising directions for future studies. Details on inference and parameter estimation, procedures for postprocessing nonnegative factor models and computing spatial importance scores, and data preprocessing are provided in the Methods section.

\section{Results}
\subsection{Factor models for spatial count data}

\begin{table}[!htb]
\centering
\small
\caption{Summary of probabilistic factor models for high-dimensional spatial count data. An X in the nonnegative, spatial, or nonspatial column indicates whether the model includes that type of latent factors. Likelihoods are listed with the default choice of each model first. gau: Gaussian or normal distribution, poi: Poisson, nb: negative binomial.}
\begin{adjustwidth}{-.25in}{0in}
\begin{tabular}{| l | l | c | c | c | l |}
\hline
\textbf{abbrev} & \textbf{model} & \textbf{nonnegative} & \textbf{spatial} & \textbf{nonspatial} & \textbf{likelihoods} \\
\hline
FA & factor analysis &  &  & X & gau \\
\hline
PNMF & probabilistic nonnegative matrix factorization & X & & X & poi, nb, gau \\
\hline
MEFISTO & MEFISTO &  & X &  & gau, poi \\
\hline
RSF & real-valued spatial factorization &  & X &  & gau \\
\hline
NSF & nonnegative spatial factorization & X & X &  & poi, nb, gau \\
\hline
NSFH & nonnegative spatial factorization hybrid & X & X & X & poi, nb, gau \\
\hline
\end{tabular}
\label{table:models}
\end{adjustwidth}
\end{table}

The data consist of a multivariate outcome $Y\in\mathbb{R}^{N\times J}$ and spatial coordinates $X\in\mathbb{R}^{N\times D}$. Let $i=1,\ldots,N$ index the observations (e.g., cells, spots, or locations with a single $(x,y)$ coordinate value), $j=1,\ldots,J$ index the outcome features (e.g., genes), and $d=1\ldots,D$ index the spatial input dimensions.

\subsubsection{Nonspatial models}
In unsupervised dimension reduction such as PCA or NMF, the goal is to represent $Y$ (or a normalized version $\tilde{Y}$) as the product of two low-rank matrices $Y\approx FW'$, where the factors matrix $F$ has dimension $N\times L$ and the loadings matrix $W$ has dimension $J\times L$, with $L\ll J$. Let $l=1,\ldots,L$ index over the components.
A probabilistic version of PCA is \textbf{factor analysis (FA)}:
\begin{align*}
\tilde{y}_{ij} &\sim \mathcal{N}(\mu_{ij},~\sigma_j^2)\\
\mu_{ij} &= \sum_{l=1}^L w_{jl} f_{il}\\
f_{il}&\sim\mathcal{N}(m_l,~s^2_l).
\end{align*}
A probabilistic version of NMF is \textbf{probabilistic nonnegative matrix factorization (PNMF)}:
\begin{align*}
y_{ij} &\sim Poi(\nu_i\lambda_{ij})\\
\lambda_{ij} &= \sum_{l=1}^L w_{jl} e^{f_{il}}\\
f_{il}&\sim\mathcal{N}(m_l,~s^2_l),
\end{align*}
where $\nu_i$ indicates a fixed size factor to account for differences in total counts per observation.
In both of these unsupervised models, the prior on the factors $f_{il}$ assumes each observation is an independent draw and ignores spatial information $\vec{x}_i$.

\subsubsection{Spatial process factorization}
In spatial process factorization, we assume that spatially adjacent observations should have correlated outcomes. We encode this assumption via a Gaussian process (GP) prior over the factors. We define \textbf{real-valued spatial factorization (RSF)} as
\begin{align*}
\tilde{y}_{ij} &\sim \mathcal{N}(\mu_{ij},~\sigma_j^2)\\
\mu_{ij} &= \sum_{l=1}^L w_{jl} f_{il}\\
f_{il} &= f_l(\vec{x}_i)\sim GP\big(\mu_l(\vec{x}_i),~k_l(\vec{x}_i,X)\big),
\end{align*}
where $\mu_l(\cdot)$ indicates a parametric mean function and $k_l(\cdot,\cdot)$ a positive semidefinite covariance (kernel) function. In our implementation, we specify the mean function as a linear function of the spatial coordinates,
\[\mu_l(\vec{x}_i) = \beta_{0l}+\vec{x}_i'\vec{\beta}_{1l}.\]
For the covariance function, we choose a Matérn kernel with fixed smoothness parameter $3/2$. We allow each component $l$ to have its own amplitude and length-scale parameters that we estimate from data. RSF is a spatial analog to factor analysis. MEFISTO has the same structure as RSF, but uses a squared exponential kernel instead of Matérn, and further places a sparsity-promoting prior on the loading weights $w_{jl}$. Our implementation is modular and can accept any positive semidefinite kernel. However, we found the Matérn kernel to have better numerical stability than the squared exponential in our experiments.

\textbf{Nonnegative spatial factorization (NSF)} is a spatial analog of probabilistic NMF (PNMF).
\begin{align*}
y_{ij}&\sim Poi(\nu_i\lambda_{ij})\\
\lambda_{ij} &= \sum_{l=1}^L w_{jl} e^{f_{il}}\\
f_{il} &= f_l(\vec{x}_i)\sim GP\big(\mu_l(\vec{x}_i),~k_l(\vec{x}_i,X)\big).
\end{align*}
For NSF, we use the same mean and kernel functions as RSF, but we additionally constrain the weights $w_{jl}\geq 0$.

We sought to quantify the relative importance of spatial versus nonspatial variation by combining NSF and PNMF into a semisupervised framework we refer to as the \textbf{nonnegative spatial factorization hybrid model (NSFH)}. NSFH consists of $L$ total factors, $T\leq L$ of which are spatial and $L-T$ are nonspatial. We recover NSF and PNMF as special cases when $T=L$ or $T=0$, respectively. By default, we set $T=L/2$.
\begin{align*}
y_{ij}&\sim Poi(\nu_i\lambda_{ij})\\
\lambda_{ij} &= \sum_{l=1}^T w_{jl}e^{f_{il}} + \sum_{l=T+1}^L v_{jl}e^{h_{il}}\\
f_{il} &= f_l(\vec{x}_i)\sim GP\big(\mu_l(\vec{x}_i),~k_l(\vec{x}_i,X)\big)\\
h_{il}&\sim\mathcal{N}(m_l,~s^2_l).
\end{align*}
Our implementations of PNMF, NSF, and NSFH are modular with respect to the likelihood, so that the negative binomial or Gaussian distributions can be substituted for the Poisson. However, in our experiments we use the Poisson data likelihood.

\subsubsection{Postprocessing nonnegative factor models}
We postprocess fitted nonnegative models (PNMF, NSF, and NSFH) by projecting factors and loadings onto a simplex. This highlights features (genes) that are enriched in particular components rather than those with high expression across all components. In the NSFH model, we interpret the ratio of loadings weights for each feature across all spatial components as a spatial importance score. This is analogous to the proportion of variance explained in PCA. In particular, a score of $1$ means that variation in a gene's expression profile across all observations is completely captured by the spatial factors, whereas a $0$ means that expression variation is completely captured by nonspatial factors. The gene-level scores can be used to identify spatially variable genes as pioneered by spatialDE \citep{svensson_spatialde_2018}. We also compute observation-level scores by switching the role of the factors and loadings matrices; details are provided in the Methods.

\subsection{Simulations: Nonnegative factorizations identify parts-based representation}

\begin{figure}[!htb]
\centering
\subfloat[ground truth]{
  \includegraphics[width=.49\linewidth]{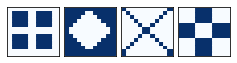}
}
\subfloat[simulated count data]{
  \includegraphics[width=.49\linewidth]{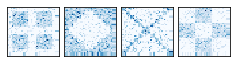}
}
\hspace{0em}
\subfloat[factor analysis]{
  \includegraphics[width=.49\linewidth]{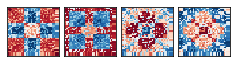}
}
\subfloat[probabilistic nonnegative matrix factorization]{
  \includegraphics[width=.49\linewidth]{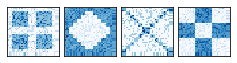}
}
\hspace{0em}
\subfloat[MEFISTO]{
  \includegraphics[width=.49\linewidth]{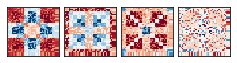}
}
\subfloat[real-valued spatial factorization]{
  \includegraphics[width=.49\linewidth]{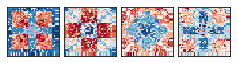}
}
\hspace{0em}
\subfloat[nonnegative spatial factorization]{
  \includegraphics[width=.49\linewidth]{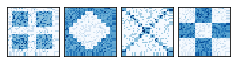}
}
\subfloat[nonnegative spatial factorization hybrid model]{
  \includegraphics[width=.49\linewidth]{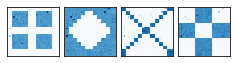}
}
\caption{\small Nonnegative factorizations recover a parts-based representation in ``quilt'' simulated multivariate spatial count data. (a) Each of $500$ features was randomly assigned to one of four nonnegative spatial factors. (b) Negative binomial count data used for model fitting. (c) Real-valued factors learned from unsupervised (nonspatial) dimension reduction. (d) as (c) but using nonnegative components. (e) Real-valued, spatially-aware factors with squared exponential kernel. (f) as (e) but with Matérn kernel. (g) Nonnegative, spatially-aware factors. (h) as (g) but with additional three nonspatial factors.}
\label{fig:quilt}
\end{figure}

To illustrate the ability of nonnegative models to recover a parts-based factorization, we simulated multivariate count data from two sets of spatial patterns. The ``ggblocks'' simulation was based on the Indian buffet process (IBP) \cite{griffiths_indian_2011}. The true factors consisted of four simple shapes in different spatial regions. In the ``quilt'' simulation, we created spatial patterns that overlapped in space. For both simulations, each of the $500$ features was an independent negative binomial draw from one of the canonical patterns (Figures \ref{fig:ggblocks}a,b and \ref{fig:quilt}a,b). Real-valued models FA, MEFISTO, and RSF estimated latent factors consisting of linear combinations of the true factors (Figures \ref{fig:ggblocks}c,e,f and \ref{fig:quilt}c,e,f). Nonnegative models PNMF, NSF, and NSFH identified each pattern as a separate factor (Figures \ref{fig:ggblocks}d,g,h and \ref{fig:quilt}d,g,h). This suggests that the parts-based representation in PNMF is preserved in NSF and NSFH. 

\clearpage
\subsection{Application to spatial transcriptomics datasets}

We examined the goodness-of-fit and interpretability of nonnegative spatial factorizations on three spatial transcriptomics datasets (Table \ref{table:data}). The Slide-seqV2 hippocampus data \citep{stickels_highly_2021} consists of $36,536$ observations, each at a unique location. The XYZeq liver data \citep{lee_xyzeq_2021} consists of $2,700$ observations at $289$ unique locations. Unlike the other protocols, each observation represents a single cell, but multiple cells are assigned to the same location. In other words, each spatial location in XYZeq contains multiple distinguishable observations, whereas in the other protocols each spatial location contains a single observation. Finally, the 10x Visium mouse brain data consists of $2,487$ observations from an anterior sagittal section, each at a unique location.

Each protocol represents a different trade-off between field-of-view (FOV) and spatial resolution. Slide-seqV2 has the smallest FOV and finest resolution, while XYZeq has the largest FOV and the coarsest resolution. Visium is intermediate in both criteria, capturing more spatial locations than XYZeq, but sacrificing single-cell resolution with each observation representing an average of multiple nearby cells.

\begin{table}[!htb]
\centering
\small
\caption{Spatial transcriptomics datasets. Slide-seqV2 and XYZeq provide single-cell resolution, whereas each Visium observation is an average of multiple cells. XYZeq combines multiple observations at each spatial location. obs: number of observations, resolution: center-to-center distance between spatial locations, FOV: field of view area.}
\begin{adjustwidth}{-0in}{0in}
\begin{tabular}{|l|l|l|l|r|r|r|r|}
\hline
\textbf{first author} & \textbf{year} & \textbf{tissue} & \textbf{protocol} & \textbf{obs} & \textbf{locations} & \textbf{resolution} & \textbf{FOV} \\
\hline
Stickels  & 2021 & hippocampus & Slide-seqV2 & $36,536$  & $36,536$   & \SI{10}{\micro\metre} & \SI{7.4}{\milli\metre\squared} \\
\hline
Lee & 2021 & liver/ tumor & XYZeq  & $2,700$ & $289$  & \SI{500}{\micro\metre}  & \SI{87.6}{\milli\metre\squared} \\
\hline
10x Genomics & 2020 & brain- anterior & Visium & $2,487$ & $2,487$ & \SI{100}{\micro\metre} & \SI{42.3}{\milli\metre\squared} \\
\hline
\end{tabular}
\label{table:data}
\end{adjustwidth}
\end{table}

To assess the utility of nonnegative and spatial factors in describing spatial sequencing data, we systematically compared all (Table~\ref{table:models}) on all three datasets. We split each dataset randomly into a training set ($95\%$ of observations) and validation set ($5\%$ of observations), and we fit each model with varying numbers of components. We quantified goodness-of-fit using Poisson deviance between the observed counts in the validation data and the predicted mean values from each model fit to the training data; a small deviance indicated that the model fit the data well.

\subsubsection{Slide-seqV2 hippocampus data}

\begin{figure}[!htb]
\centering
\subfloat[Out-of-sample generalization error]{
  \includegraphics[width=0.4\linewidth]{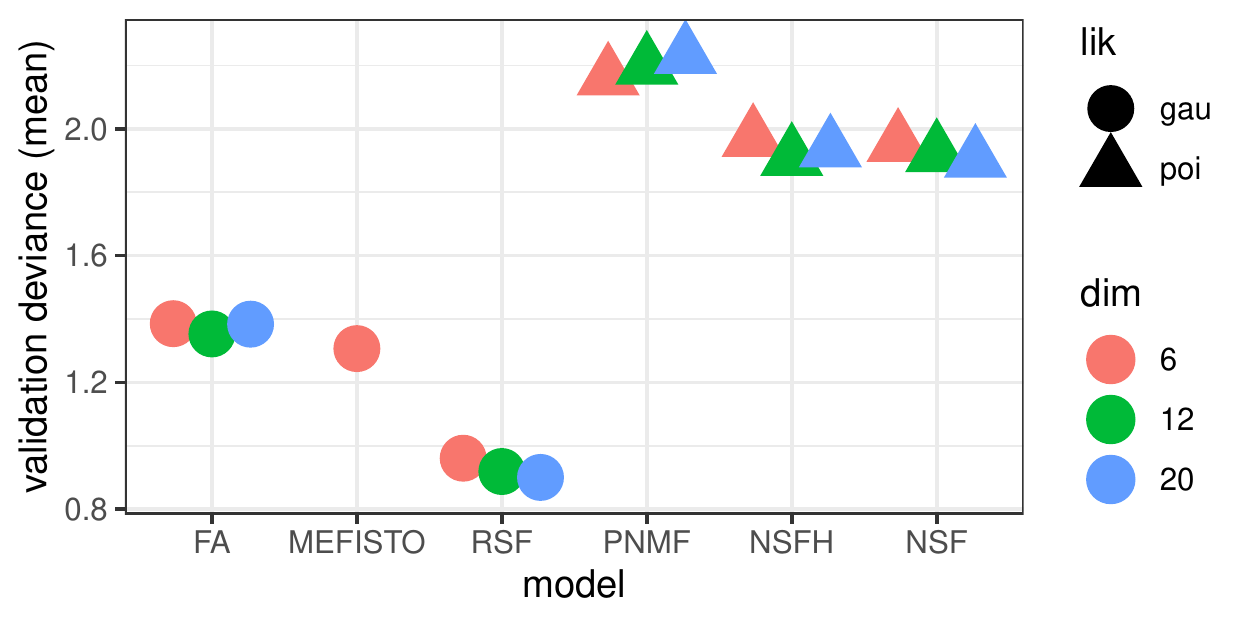}
}
\subfloat[Feature spatial importance]{
  \includegraphics[width=.28\linewidth]{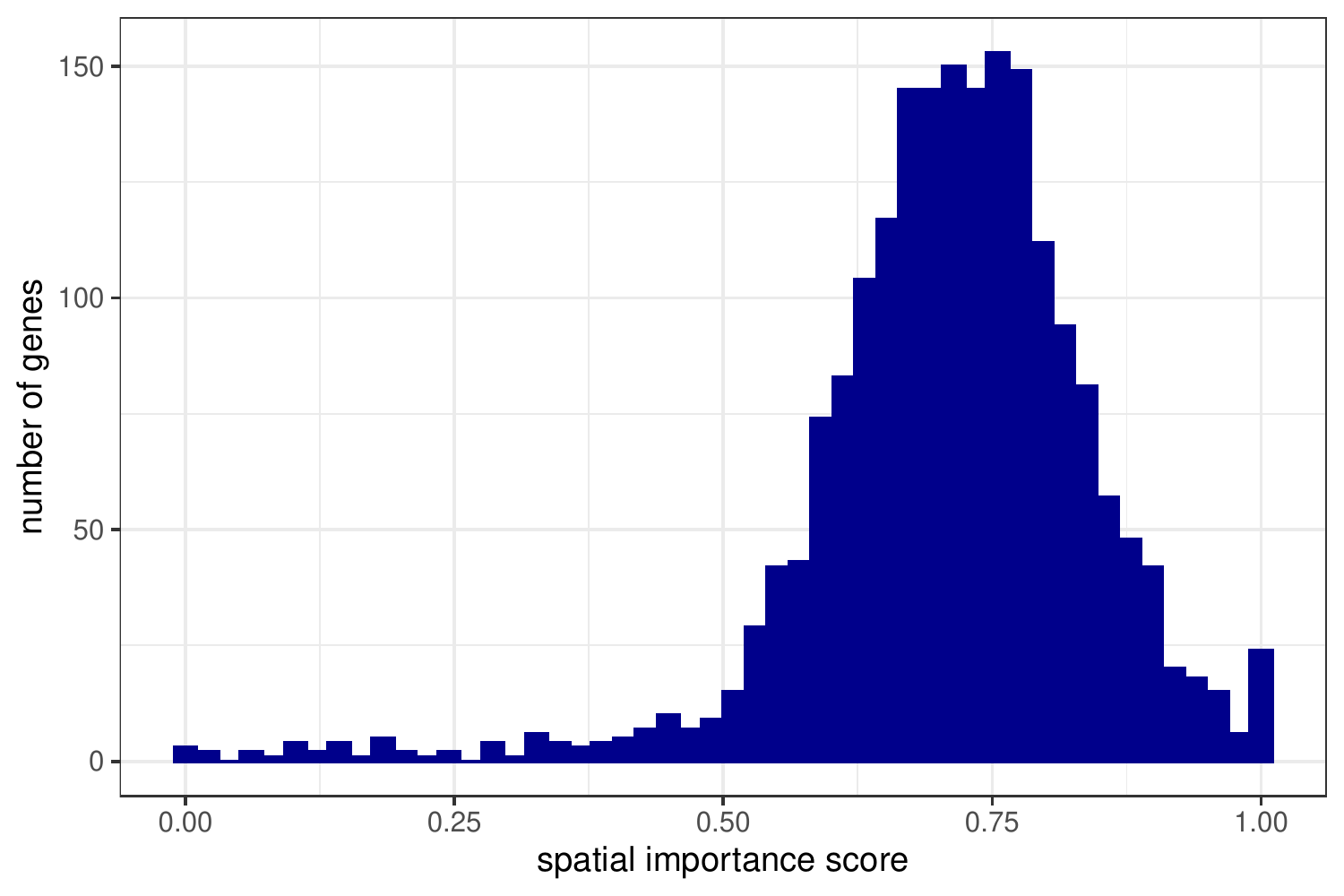}
}
\subfloat[Observation spatial importance]{
  \includegraphics[width=.28\linewidth]{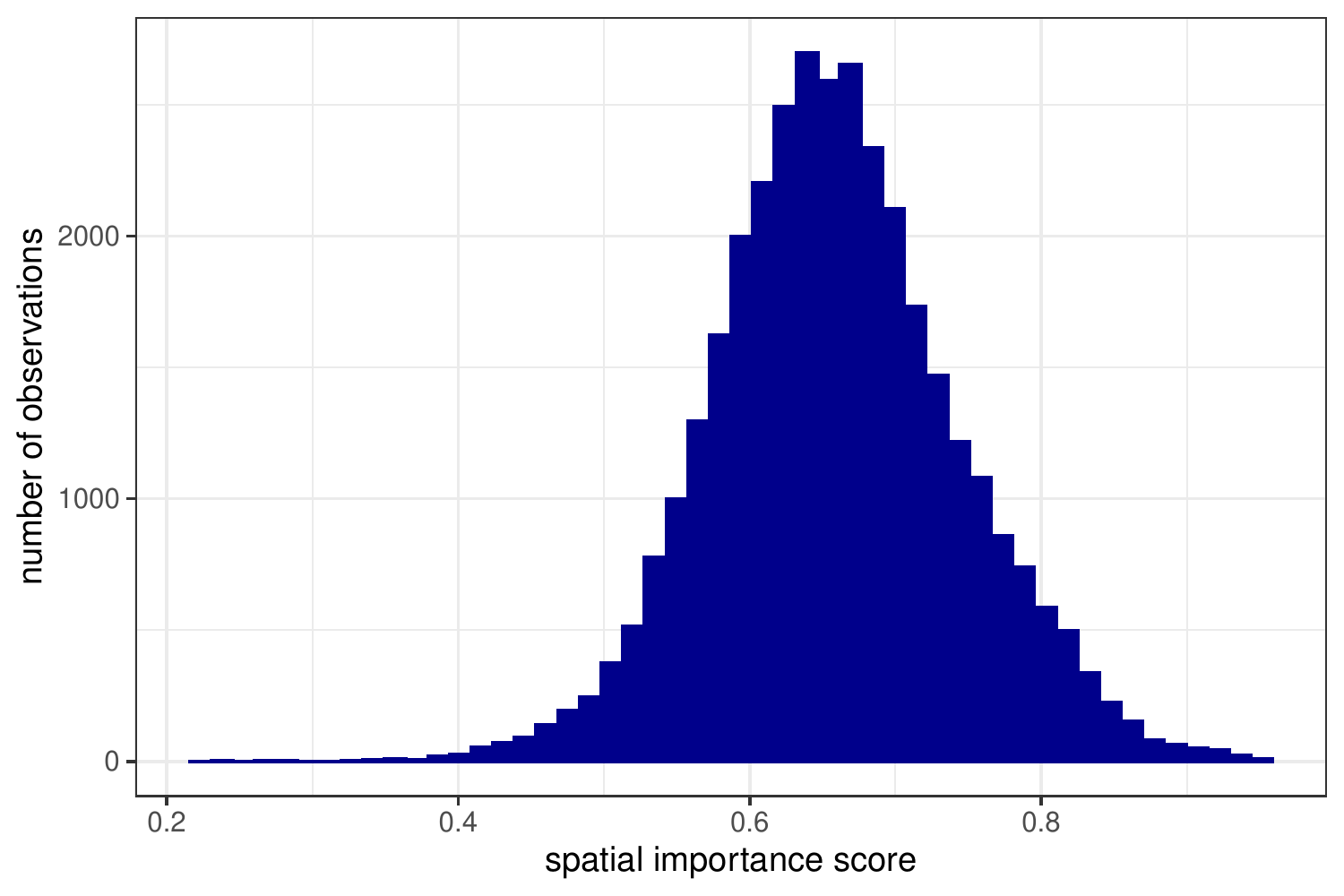}
}
\caption{\small Benchmarking spatial and nonspatial factor models on Slide-seqV2 mouse hippocampus gene expression data. (a) Poisson deviance on held-out validation data. Lower deviance indicates better generalization accuracy. All spatial models used $2,000$ inducing points. lik: likelihood, dim: number of latent dimensions (components), FA: factor analysis, RSF: real-valued spatial factorization, PNMF: probabilistic nonnegative matrix factorization, NSF: nonnegative spatial factorization, NSFH: NSF hybrid model. (b) Each feature (gene) was assigned a spatial importance score derived from NSFH fit with $20$ components ($10$ spatial and $10$ nonspatial). A score of $1$ indicates spatial components explain all of the variation. (c) as (b) but with observations instead of features.}
\label{fig:sshippo-gof-val}
\end{figure}

On the Slide-seqV2 mouse hippocampus dataset, real-valued factor models had lower validation deviance (higher generalization accuracy) than nonnegative models (Figure \ref{fig:sshippo-gof-val}a). This was to be expected since real-valued factors can encode more information than nonnegative factors. The unsupervised models (FA and PNMF) had higher deviance than their spatially-aware analogs. Surprisingly, RSF outperformed MEFISTO despite having nearly the same probabilistic structure. We attribute this difference to the choice of spatial covariance function \citep{stephenson_measuring_2021}-- MEFISTO uses a squared exponential kernel whereas RSF uses a Matérn($3/2$) kernel. Matérn kernels produce spatial functions that are less smooth, which may better accommodate sharp transitions between adjacent biological tissue layers such as those of the brain. We were unable to fit MEFISTO models with more than six components because they ran out of memory.

In terms of sparsity, MEFISTO had the highest fraction of zero entries in the loadings matrix due to its sparsity promoting prior, followed by the nonnegative models NSFH, PNMF, and NSF (Figure \ref{fig:sshippo-sparsity-timing}a). Increasing the number of components also increased the sparsity. The time to convergence was comparable for all spatial models, with nonspatial models converging substantially faster (Figure \ref{fig:sshippo-sparsity-timing}b). Among nonnegative models, the negative binomial likelihood took longer to converge but did not reduce generalization error (Figure \ref{fig:sshippo-sparsity-timing}c-d). Both NSF and NSFH had similar deviances, suggesting that including a mixture of spatial and nonspatial components (NSFH) did not degrade generalization in comparison to a strictly spatial model (NSF).

We examined the biological relevance of nonnegative factorization by focusing on the NSFH model with $M=3000$ inducing points and $L=20$ components ($10$ spatial and $10$ nonspatial). Each factor was summarized by its (variational) approximate posterior mean. For each spatial factor this is a function in the $(x,y)$ spatial coordinate system. For each nonspatial factor the posterior is a vector with one value per observation. Spatial importance scores indicate that most genes are strongly spatial, although a small number are entirely nonspatial (Figure \ref{fig:sshippo-gof-val}b). At the observation level, spatial scores were less extreme, suggesting that both spatial and nonspatial factors are needed to explain gene expression at each location (Figure \ref{fig:sshippo-gof-val}c).

\begin{figure}[!htb]
\centering
\subfloat[NSFH spatial factors]{
  \includegraphics{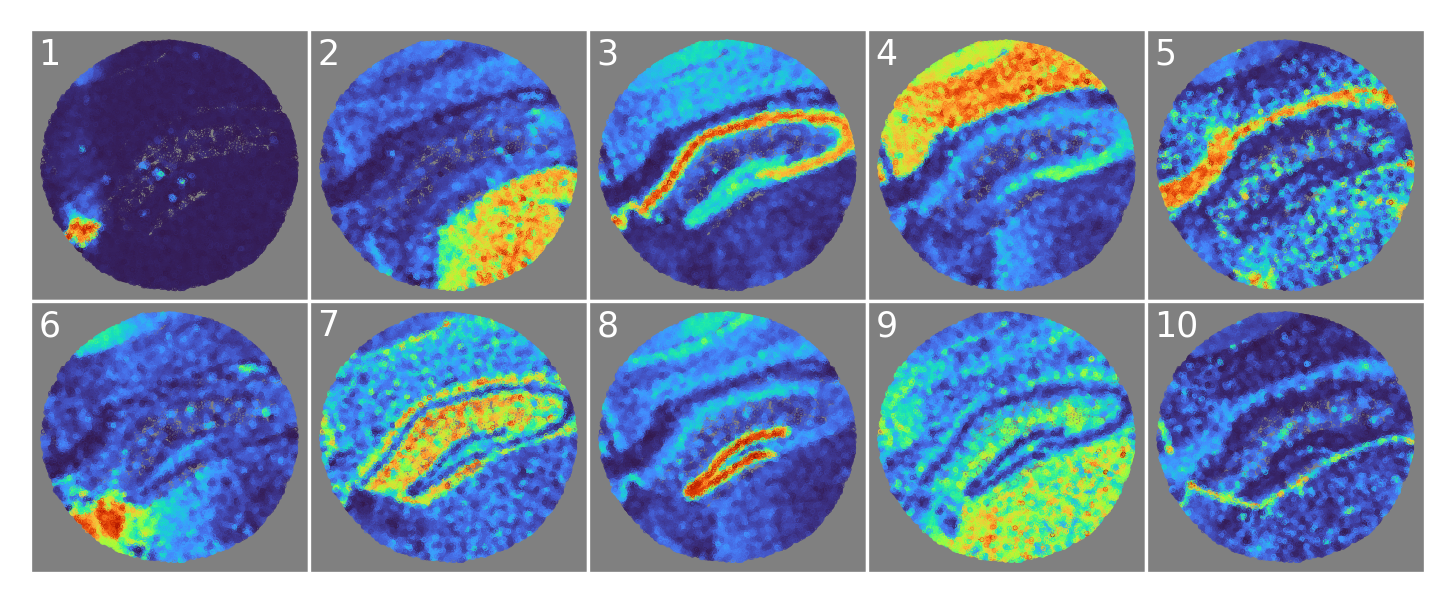}
}
\hspace{0em}
\subfloat[NSFH spatially variable genes]{
  \includegraphics{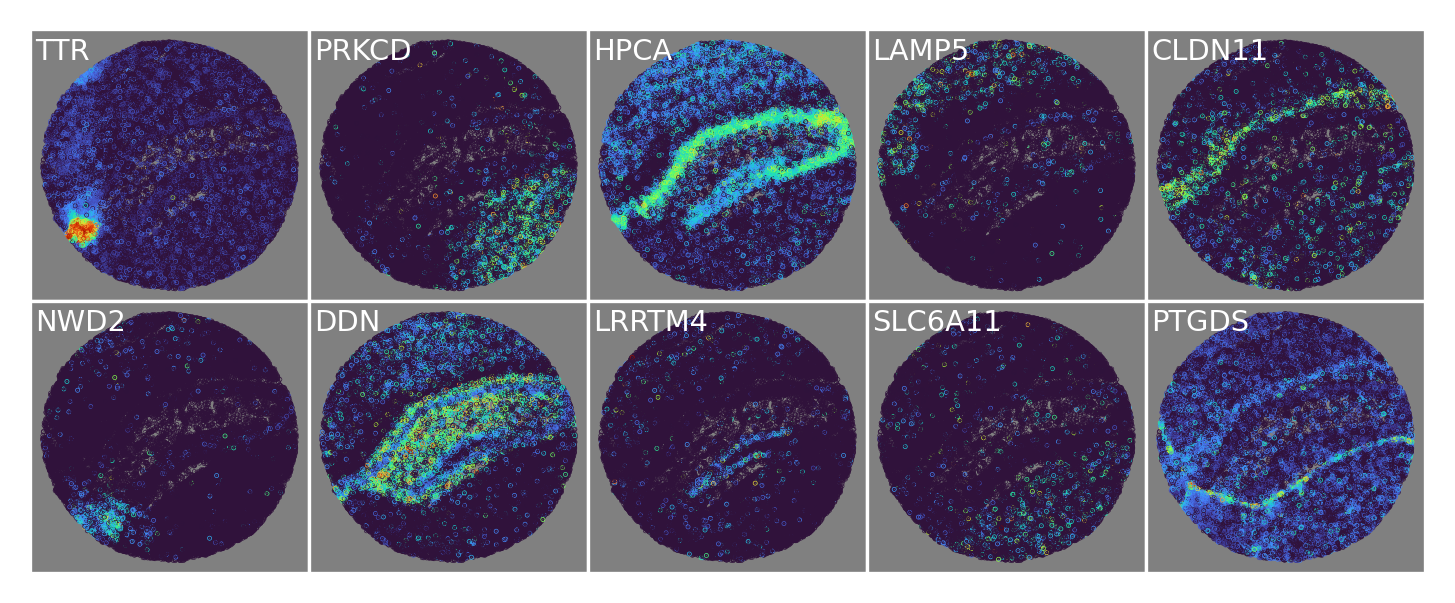}
}
\hspace{0em}
\subfloat[NSFH nonspatial factors]{
  \includegraphics{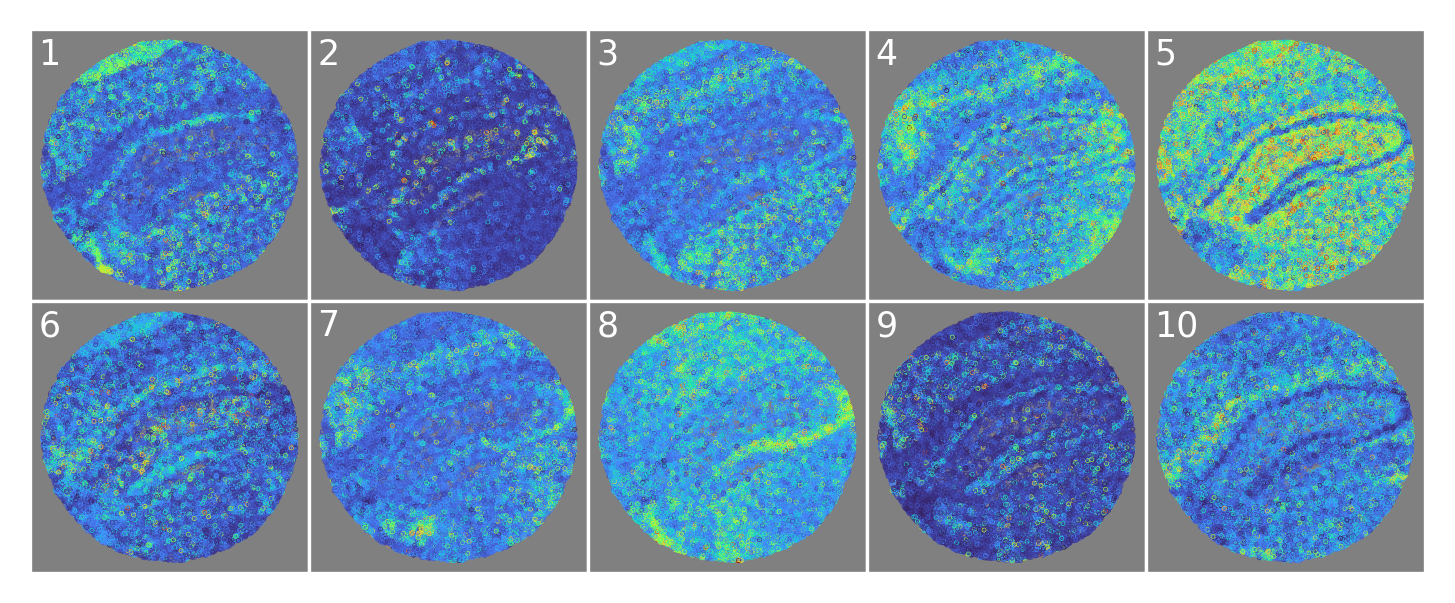}
}
\caption{\small Nonnegative spatial factorization hybrid model (NSFH) combines spatial and nonspatial factors in Slide-seqV2 mouse hippocampus gene expression data. Field-of-view is a coronal section with left indicating the medial direction and right the lateral direction. (a) Heatmap (red=high, blue=low) of square-root transformed posterior mean of $10$ spatial factors mapped into the $(x,y)$ coordinate space. (b) as (a) but mapping expression levels of top genes with strongest enrichment to each spatial component. (c) as (a) but mapping $10$ nonspatial factors from the same model.}
\label{fig:sshippo-npfh20-heatmap}
\end{figure}

Spatial factors mapped to specific brain regions (Figure \ref{fig:sshippo-npfh20-heatmap}a) such as the choroid plexus (1), medial habenula (6), and dentate gyrus (8). Even the thin meninges layer was distinguishable (10), underscoring the high spatial resolution of the Slide-seqV2 protocol. We identified genes with the highest enrichment to individual components by examining the loadings matrix. Spatial gene expression patterns mirrored the spatial factors to which they were most associated (Figure \ref{fig:sshippo-npfh20-heatmap}b). Nonspatial factors were generally dispersed across the field of view (Figure \ref{fig:sshippo-npfh20-heatmap}c). Finally, we used the top genes for each component to identify cell types and biological processes (Table \ref{table:sshippo-ctypes}) using scfind\footnote{\url{https://scfind.sanger.ac.uk/}} \citep{lee_fast_2021} and the Panglao database\footnote{\url{https://panglaodb.se}} \citep{franzen_panglaodb_2019}. For example, spatial component 5 identified the corpus callosum, a white-matter region where myelination is crucial. Similarly, the top cell type for spatial component 10 capturing the meninges layer was meningeal cells. Generally, neurons and glia were the most common cell types across all components.

\begin{table}[!htb]
\centering
\small
\caption{Nonnegative spatial factorization hybrid model (NSFH) identifies biologically distinct components in Slide-seqV2 mouse hippocampus.}
\begin{adjustwidth}{-1in}{0in}
\begin{tabular}{|r|l|p{0.17\linewidth}|p{0.15\linewidth}|p{0.25\linewidth}|p{0.35\linewidth}|}
\hline
\textbf{dim} & \textbf{type} & \textbf{brain regions} & \textbf{cell types} & \textbf{genes} & \textbf{GO biological processes} \\ \hline
1 & spat & choroid plexus of third ventricle & Choroid plexus cells & \textit{TTR, ENPP2, IFI27, TRPM3, STK39} & T cell migration, cellular response to chemokine \\ \hline
2 & spat & thalamus & Interneurons & \textit{PRKCD, TNNT1, RAMP3, NTNG1, PDP1} & regulation of presynaptic cytosolic calcium ion concentration,   proteoglycan metabolic process \\ \hline
3 & spat & CA1-3 (Ammon's Horn) pyramidal layer & Neurons & \textit{HPCA, NEUROD6, CRYM, WIPF3, CPNE6} & positive regulation of dendritic spine morphogenesis, postsynaptic   modulation of chemical synaptic transmission \\ \hline
4 & spat & cerebral cortex & Neurons & LAMP5, \textit{3110035E14RIK, STX1A, MEF2C, EGR1} & myeloid leukocyte differentiation, hormone biosynthetic process \\ \hline
5 & spat & fiber tracts/ corpus callosum & Oligodendrocytes & \textit{CLDN11, MAL, MAG, PLP1, MOG} & myelination, central nervous system myelination \\ \hline
6 & spat & medial habenula (thalamus) & Neurons & \textit{NWD2, TAC2, CALB2, NECAB2, ZCCHC12} & steroid biosynthetic process, sex differentiation \\ \hline
7 & spat & CA strata and dentate gyrus molecular layer & Astrocytes & \textit{DDN, SLC1A3, CST3, PSD, CABP7} & learning, response to amino acid \\ \hline
8 & spat & dentate gyrus granule layer & Neurons & \textit{LRRTM4, STXBP6, SLC8A2, 2010300C02RIK, PLXNA4} & central nervous system projection neuron axonogenesis, regulation of cytoskeleton organization \\ \hline
9 & spat & multiple & Astrocytes & \textit{SLC6A11, SPARC, SLC4A4, KCNJ10, ATP1A2} & negative regulation of blood coagulation, cellular amino acid catabolic   process \\ \hline
10 & spat & meninges & Meningeal cells & \textit{PTGDS, GFAP, APOD, FABP7, FXYD1} & nitric oxide mediated signal transduction, epithelial cell proliferation \\ \hline
1 & nsp &  & Neurons & \textit{MEG3, SNHG11, MIAT, CPNE7, TTC14} & mRNA processing, RNA splicing \\ \hline
2 & nsp &  & GABAergic neurons & \textit{SST, NPY, GAD2, GAD1, CNR1} & neurotransmitter metabolic process, negative regulation of catecholamine   secretion \\ \hline
3 & nsp &  & Neurons & \textit{CHGA, RAB3C, HSPA4L, CKMT1, SYT4} & chemical synaptic transmission, regulation of short-term neuronal   synaptic plasticity \\ \hline
4 & nsp &  & GABAergic neurons & \textit{PVALB, VAMP1, CPLX1, MT-ND1, SCRT1} & mitochondrial respiratory chain complex I assembly, aerobic respiration \\ \hline
5 & nsp &  & Astrocytes & \textit{MT-RNR2, MT-RNR1, 2900052N01RIK, MAP2, MT-ND5} & electron transport coupled proton transport, response to nicotine \\ \hline
6 & nsp &  & Astrocytes & \textit{GM3764, MALAT1, GPC5, LSAMP, TRPM3} & cell adhesion, synaptic membrane adhesion \\ \hline
7 & nsp &  & Neurons & \textit{NEFM, NEFH, MAP1B, VAMP1, SLC24A2} & cell adhesion, intermediate filament cytoskeleton organization \\ \hline
8 & nsp &  & Interneurons & \textit{NPTXR, SYN2, STMN2, NCALD, YWHAH} & mitotic cell cycle, thyroid gland development \\ \hline
9 & nsp &  & Neurons & \textit{NRG3, FGF14, CSMD1, DLG2, KCNIP4} & social behavior, positive regulation of synapse assembly \\ \hline
10 & nsp &  &  & \textit{MIR6236, LARS2, CMSS1, HEXB, CAMK1D} & translation, positive regulation of signal transduction by p53 class mediator \\ \hline
\end{tabular}
\end{adjustwidth}
\label{table:sshippo-ctypes}
\end{table}

\clearpage
\subsubsection{XYZeq liver data}

On the XYZeq mouse liver dataset, real-valued factor models again had lower validation deviance than nonnegative models and spatial models again outperformed their nonspatial analogs (Figure \ref{fig:sshippo-gof-val}). The strictly spatial NSF model had slightly lower deviance than the hybrid spatial and nonspatial model NSFH.

\begin{figure}[!htb]
\centering
\subfloat[Out-of-sample generalization error]{
  \includegraphics[width=.4\linewidth]{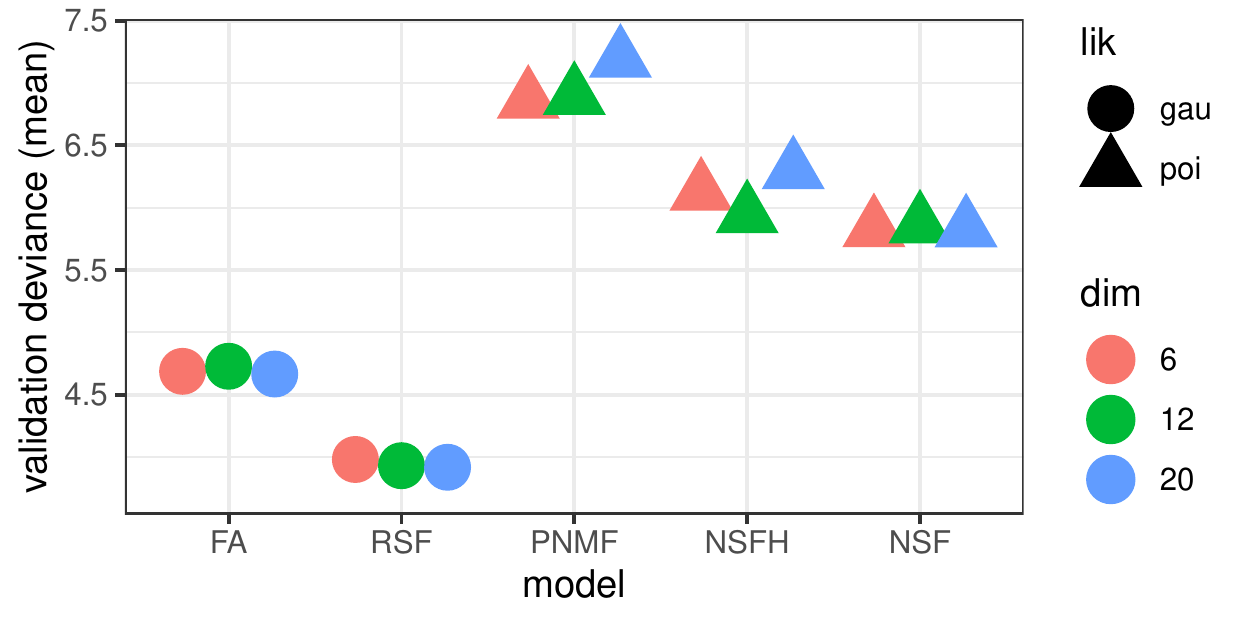}
}
\subfloat[Gene spatial importance]{
  \includegraphics[width=.28\linewidth]{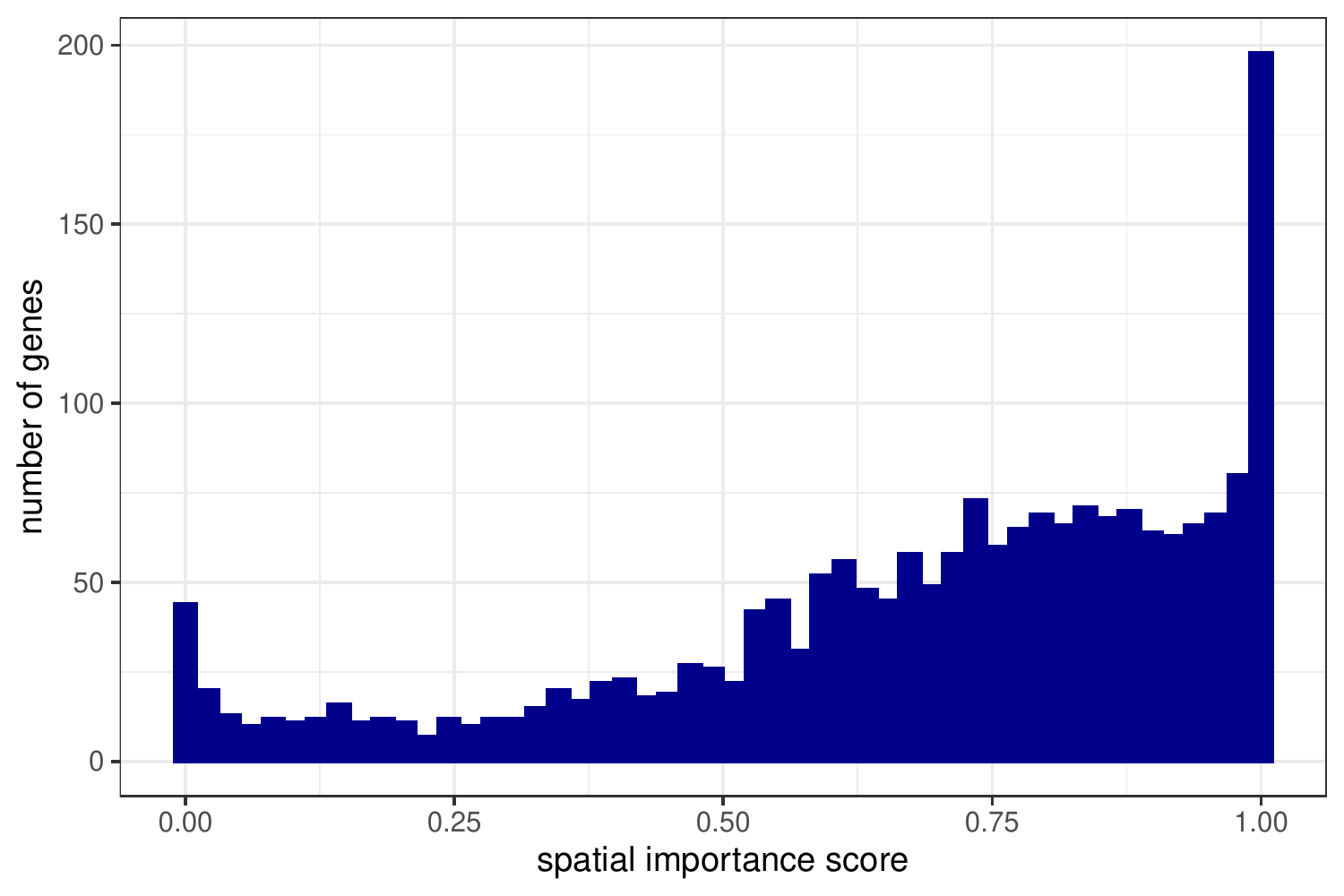}
}
\subfloat[Observation spatial importance]{
  \includegraphics[width=.28\linewidth]{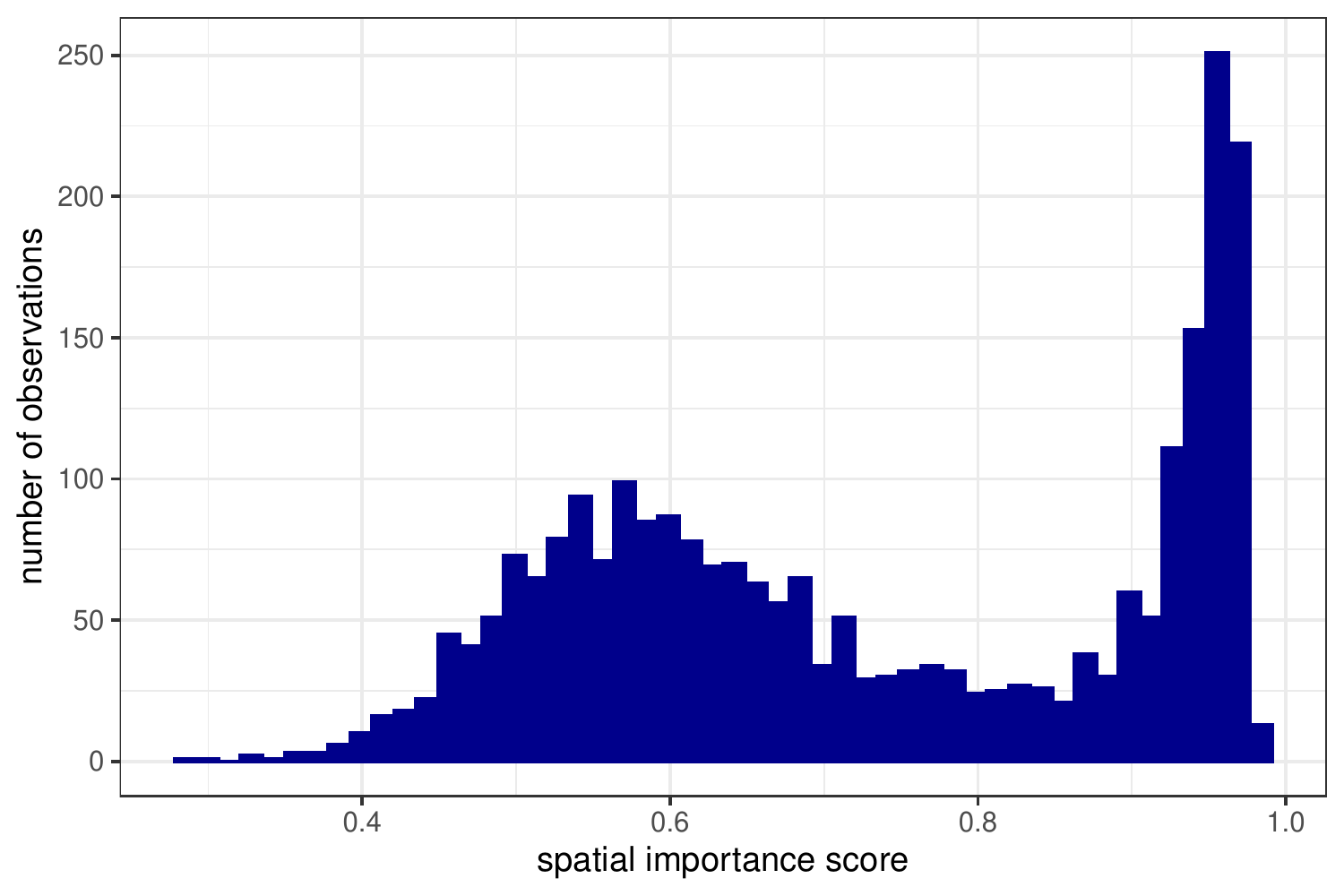}
}
\caption{\small Benchmarking spatial and nonspatial factor models on XYZeq mouse liver gene expression data. Lower deviance indicates higher generalization accuracy. All spatial models used $288$ inducing points. lik: likelihood, dim: number of latent dimensions (components), FA: factor analysis, RSF: real-valued spatial factorization, PNMF: probabilistic nonnegative matrix factorization, NSF: nonnegative spatial factorization, NSFH: NSF hybrid model. (b) Each feature (gene) was assigned a spatial importance score derived from NSFH fit with $6$ components ($3$ spatial and $3$ nonspatial). A score of $1$ indicates spatial components explain all of the variation. (c) as (b) but with observations instead of features.}
\label{fig:xyz-gof-val}
\end{figure}

\begin{figure}[!htb]
\centering
\subfloat[NSFH spatial factors]{
  \includegraphics{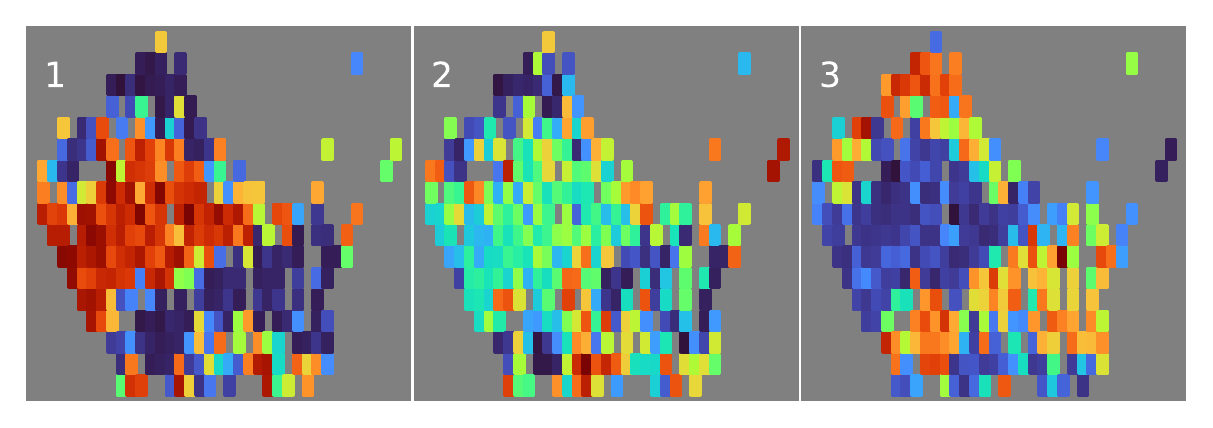}
}
\hspace{0em}
\subfloat[NSFH spatially variable genes]{
  \includegraphics{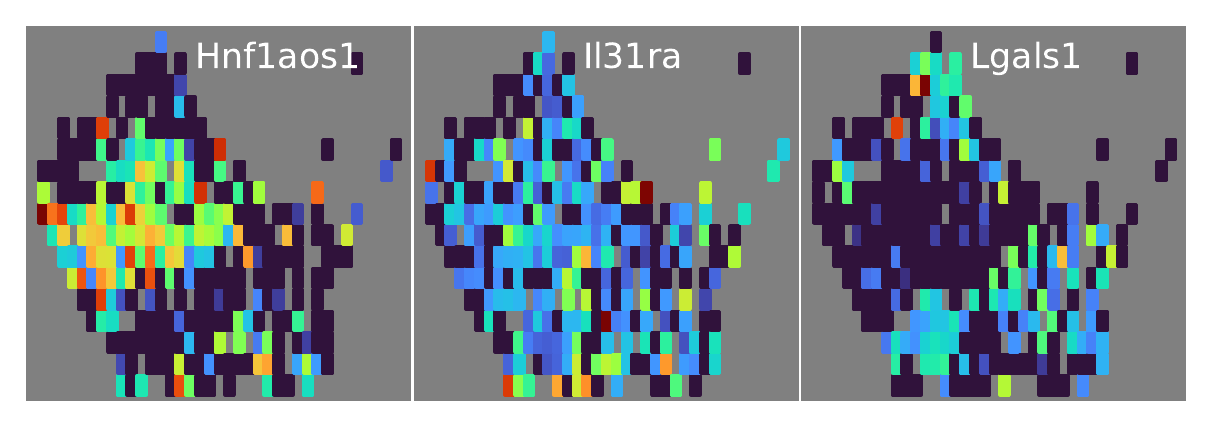}
}
\hspace{0em}
\subfloat[NSFH nonspatial factors]{
  \includegraphics{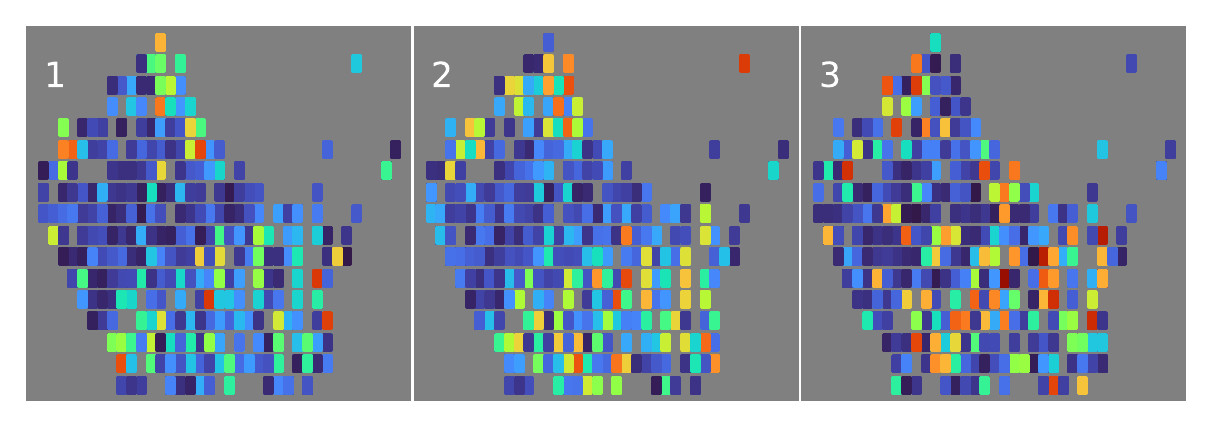}
}
\caption{\small Nonnegative spatial factorization hybrid model (NSFH) combines spatial and nonspatial factors in XYZeq mouse liver gene expression data. (a) Heatmap (red=high, blue=low) of square-root transformed posterior mean of $3$ spatial factors mapped into the $(x,y)$ coordinate space. (b) as (a) but mapping expression levels of top genes with strongest enrichment to each spatial component. (c) as (a) but mapping $3$ nonspatial factors from the same model.}
\label{fig:xyz-liv-npfh6-heatmap}
\end{figure}

Focusing on the NSFH model with $M=288$ inducing points and $L=6$ components, we found a strikingly bimodal distribution of spatial importance scores for both genes and observations (Figure \ref{fig:xyz-gof-val}b-c). However, like the Slide-seqV2 data, most scores were greater than $0.5$, suggesting spatial variation was more explanatory than nonspatial overall. The first spatial factor identified normal liver tissue while the other spatial factors were associated with the tumor regions (Figure \ref{fig:xyz-liv-npfh6-heatmap}a). Genes associated with spatial component 1 indicated an enrichment of hepatocytes, while genes in the other components were associated with immune cells (Figure \ref{fig:xyz-liv-npfh6-heatmap}b, Table \ref{table:xyz-ctypes}). The nonspatial factors again showed no distinct spatial patterns for these data (Figure \ref{fig:xyz-liv-npfh6-heatmap}c), although they were associated with particular cell types and biological processes (Table \ref{table:xyz-ctypes}).

\begin{table}[!htb]
\centering
\small
\caption{Nonnegative spatial factorization hybrid model (NSFH) identifies biologically distinct components in XYZeq mouse liver.}
\begin{adjustwidth}{-1in}{0in}
\begin{tabular}{|r|l|l|p{0.35\linewidth}|p{0.5\linewidth}|}
\hline
\textbf{dim} & \textbf{type} & \textbf{cell types} & \textbf{genes} & \textbf{GO biological processes} \\ \hline
1 & spat & Hepatocytes & \textit{HNF1AOS1, CPS1, CYP2E1, AKR1C6, MUG2} & cellular amino acid catabolic process, xenobiotic metabolic process \\ \hline
2 & spat &  & \textit{IL31RA, SEMA5A, TRPM3, PLCD1, FOXP4} & positive regulation of cholesterol esterification, inclusion body   assembly \\ \hline
3 & spat & Macrophages & \textit{LGALS1, S100A6, S100A4, RPL30, KLF6} & translation, cytoplasmic translation \\ \hline
1 & nsp & Fibroblasts & \textit{KIF26B, MEDAG, LAMA4, NGF, COL1A2} & regulation of cellular response to vascular endothelial growth factor   stimulus, collagen fibril organization \\ \hline
2 & nsp &  & \textit{HMGA2, SLC35F1, FAM19A1, TENM4, HS3ST5} & cell fate specification, specification of animal organ identity \\ \hline
3 & nsp & Macrophages & \textit{ARHGAP15, DOCK10, MYO1F, LY86, HCK} & negative regulation of leukocyte apoptotic process, negative regulation of immune response \\ \hline
\end{tabular}
\end{adjustwidth}
\label{table:xyz-ctypes}
\end{table}

\clearpage
\subsubsection{Visium brain data}

On the Visium mouse brain data, goodness-of-fit results in terms of validation deviance (lower values indicate better generalization accuracy) were markedly different than the other two datasets. First, real-valued models did not dominate nonnegative models. Both NSF and NSFH had lower deviance than factor analysis and MEFISTO. In fact, NSF had generalization accuracy comparable to the best-performing model (RSF). Furthermore, it was necessary to increase the number of components in NSFH to reduce the deviance to a level comparable with NSF, whereas in the other datasets deviance did not vary dramatically with the number of components, and NSFH had similar performance to NSF. However, consistent with the other datasets, nonspatial models generally had higher deviance than their spatial analogs, reinforcing the importance of including spatial information in out-of-sample prediction, interpolation, and generalization. Finally, the dramatic improvement in fit of RSF results over MEFISTO underscores the importance of choosing an appropriate GP kernel.

\begin{figure}[!htb]
\centering
\subfloat[Out-of-sample generalization error]{
  \includegraphics[width=.49\linewidth]{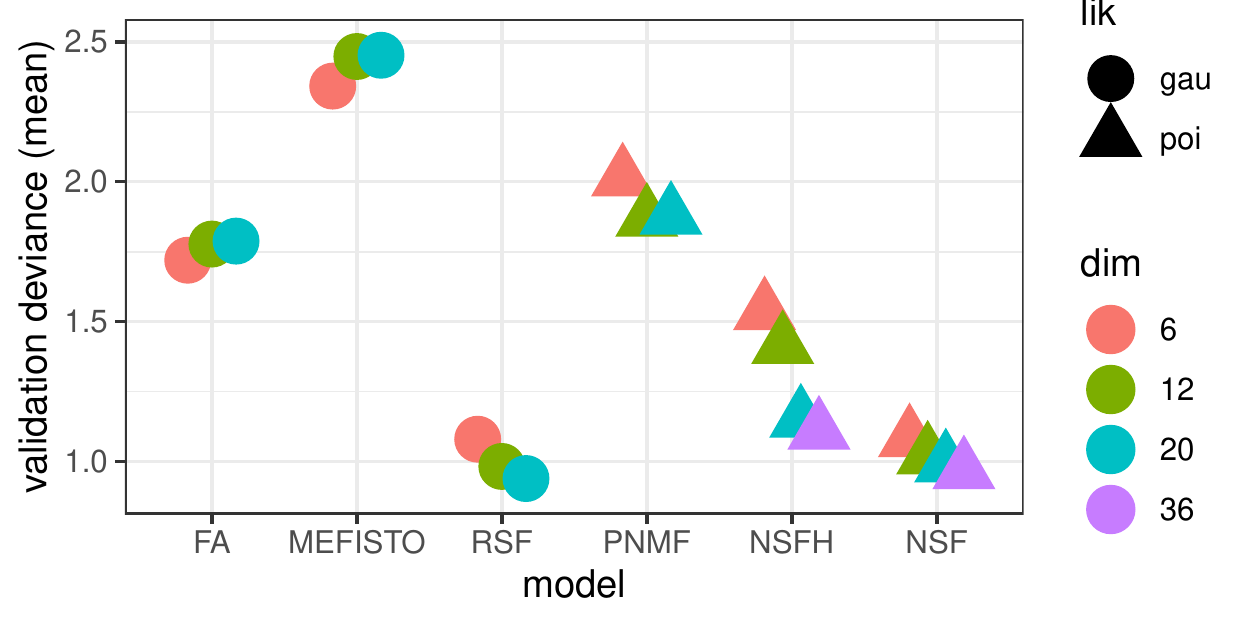}
}
\subfloat[Diagram of brain regions]{
  \includegraphics[width=.49\linewidth]{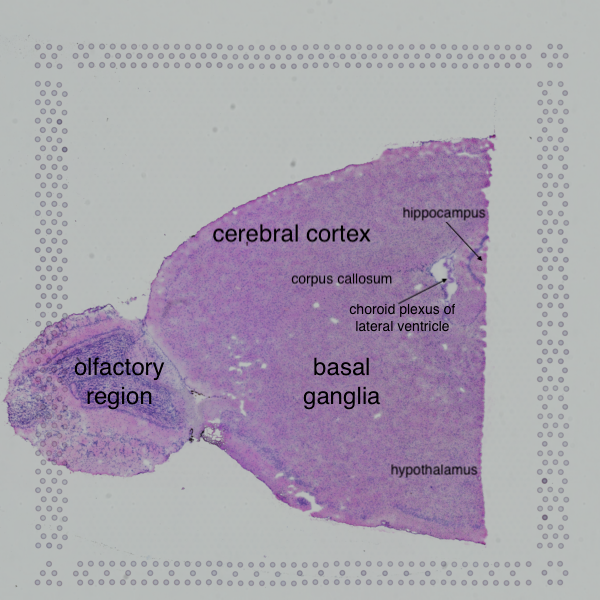}
}
\hspace{0em}
\subfloat[Gene spatial importance]{
  \includegraphics[width=.49\linewidth]{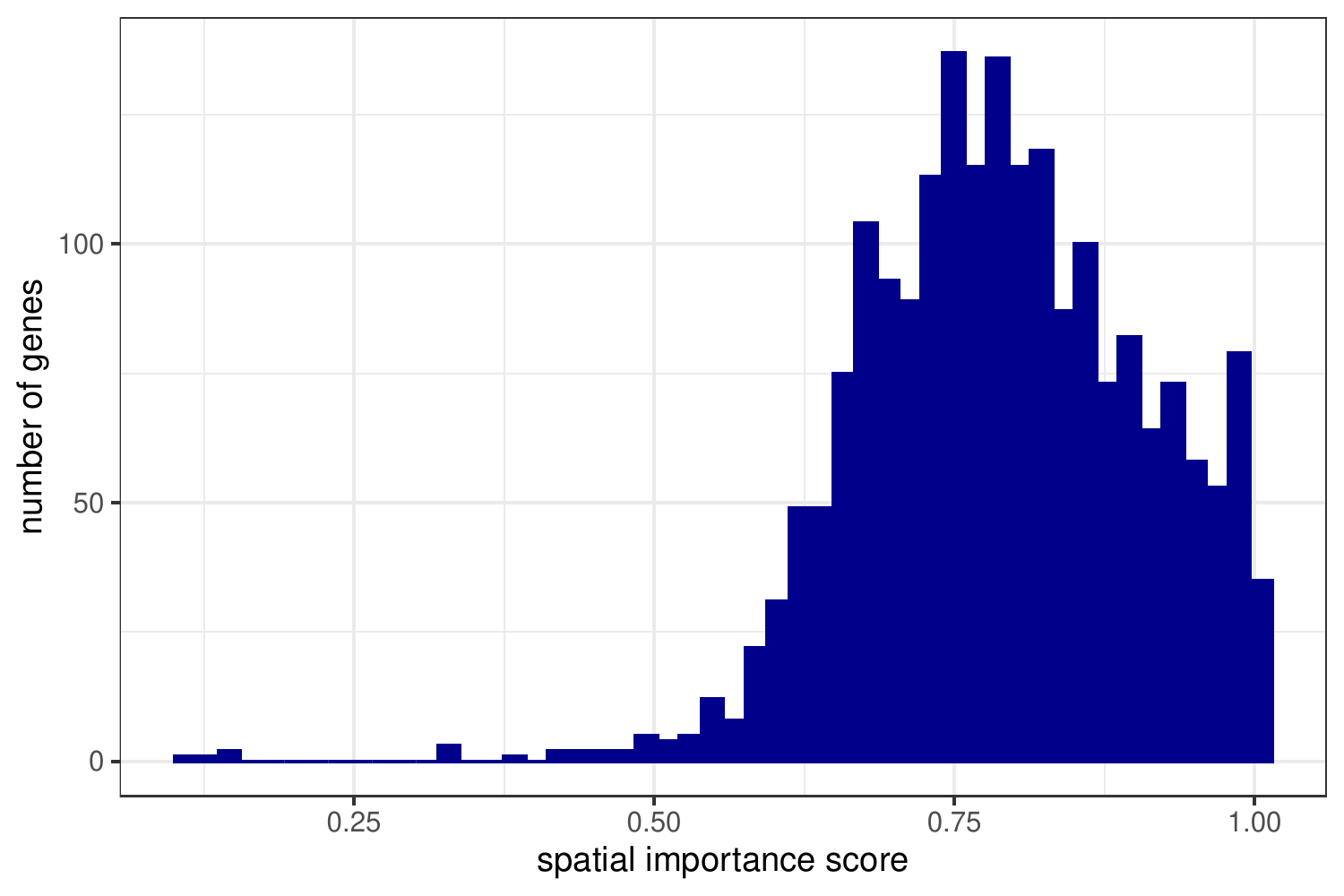}
}
\subfloat[Observation spatial importance]{
  \includegraphics[width=.49\linewidth]{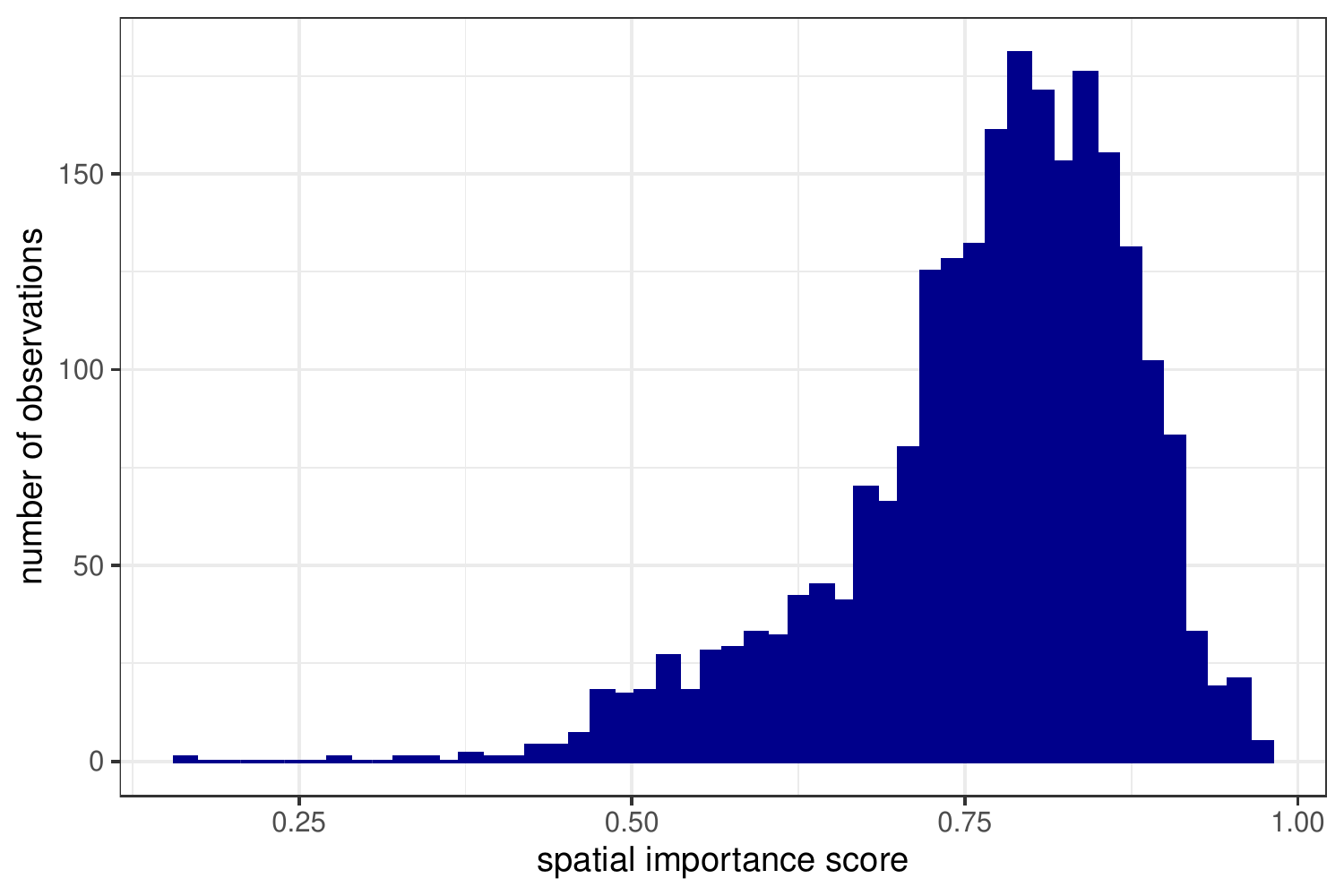}
}
\caption{\small Benchmarking spatial and nonspatial factor models on Visium mouse brain gene expression data. Lower deviance indicates better generalization accuracy. All spatial models used $2,363$ inducing points. lik: likelihood, dim: number of latent dimensions (components), FA: factor analysis, RSF: real-valued spatial factorization, PNMF: probabilistic nonnegative matrix factorization, NSF: nonnegative spatial factorization, NSFH: NSF hybrid model. (b) Diagram of major brain regions, annotation by the authors, original image credit to 10x Genomics. (c) Each feature (gene) was assigned a spatial importance score derived from NSFH fit with $20$ components ($10$ spatial and $10$ nonspatial). A score of $1$ indicates spatial components explain all of the variation. (d) as (c) but with observations instead of features.}
\label{fig:vz-gof-val}
\end{figure}

\begin{figure}[!htb]
\centering
\subfloat[NSFH spatial factors]{
  \includegraphics[width=0.9\linewidth]{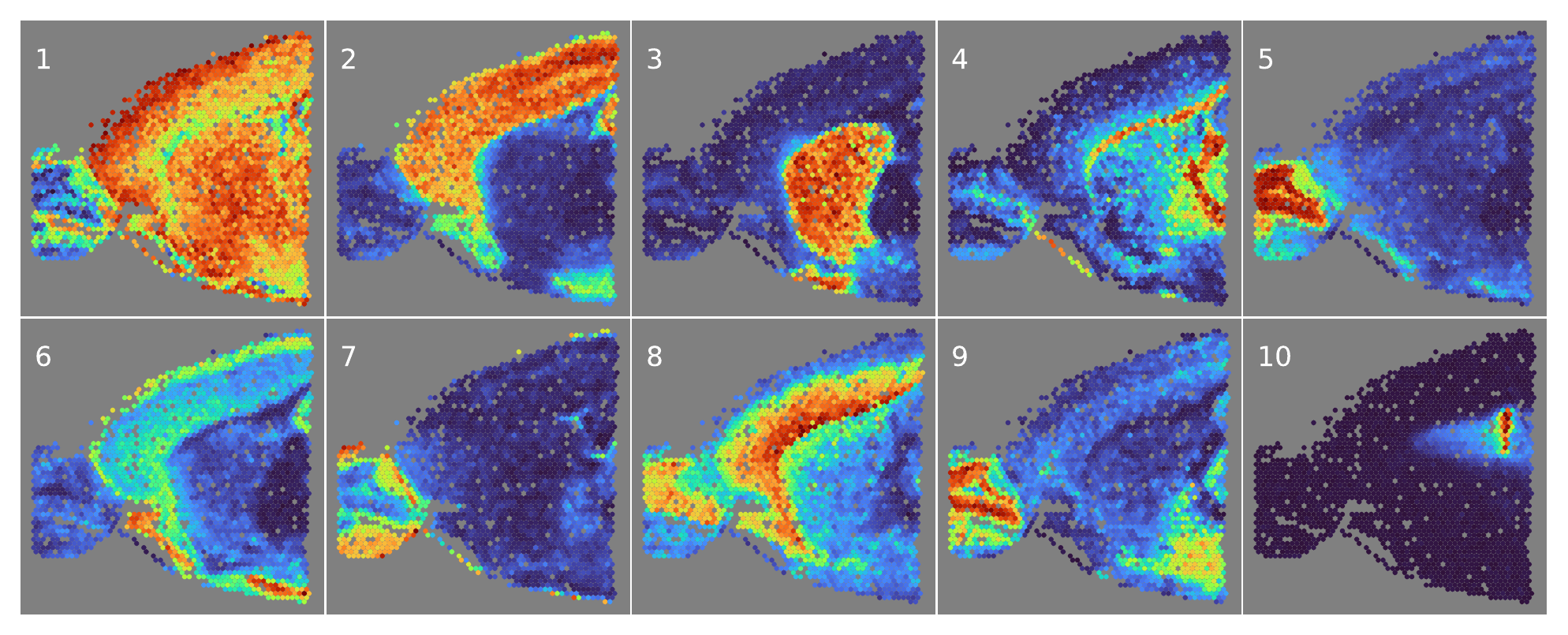}
}
\hspace{0em}
\subfloat[NSFH spatially variable genes]{
  \includegraphics[width=0.9\linewidth]{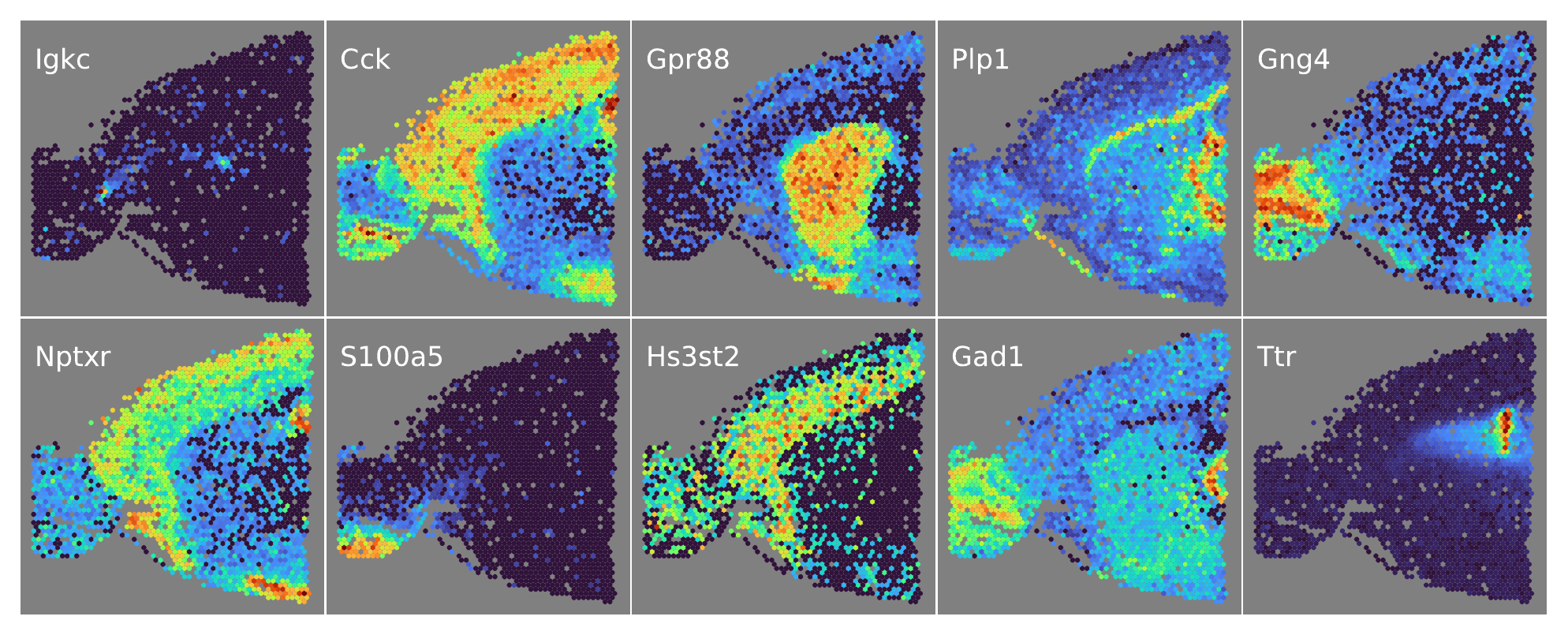}
}
\hspace{0em}
\subfloat[NSFH nonspatial factors]{
  \includegraphics[width=0.9\linewidth]{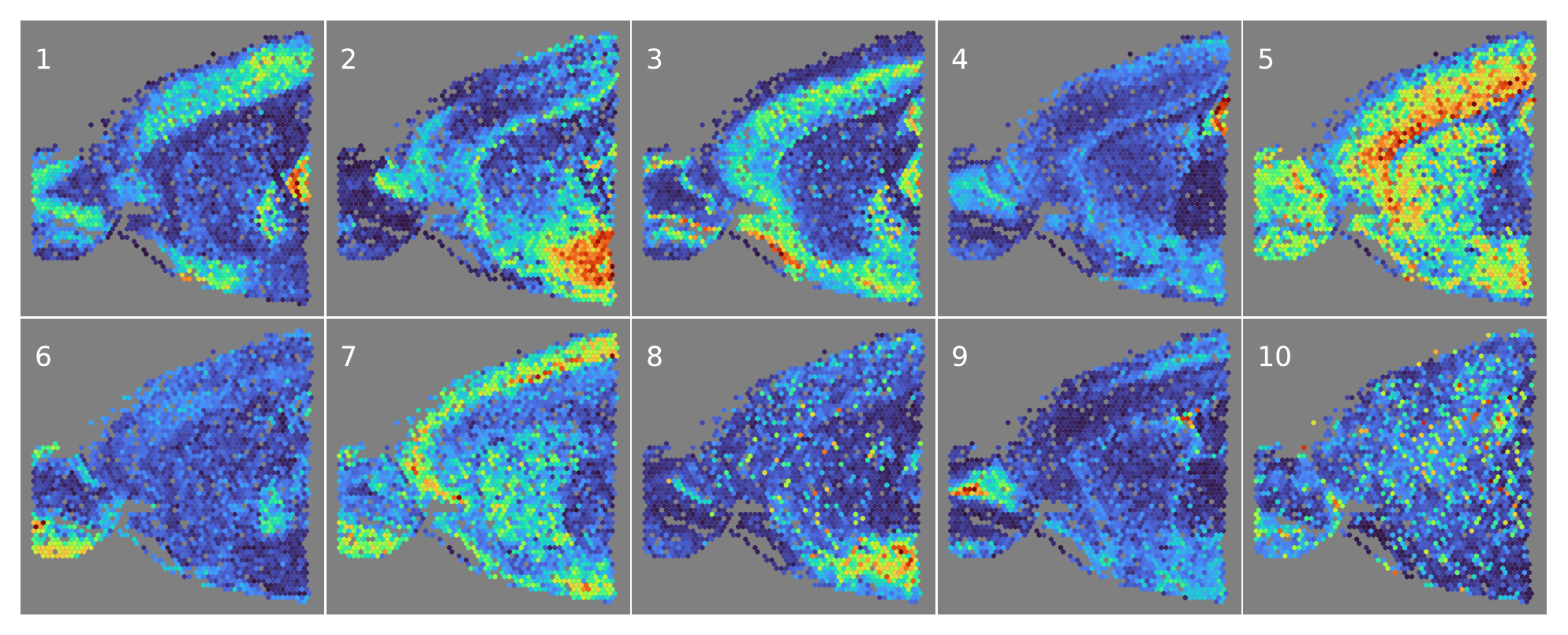}
}
\caption{\small Nonnegative spatial factorization hybrid model (NSFH) combines spatial and nonspatial factors in Visium mouse brain gene expression data. Field-of-view is a sagittal section with left indicating the anterior direction and right the posterior direction. (a) Heatmap (red=high, blue=low) of square-root transformed posterior mean of $10$ spatial factors mapped into the $(x,y)$ coordinate space. (b) as (a) but mapping expression levels of top genes with strongest enrichment to each spatial component. (c) as (a) but mapping $10$ nonspatial factors from the same model.}
\label{fig:vz-npfh20-heatmap}
\end{figure}

We next focused on interpretation of the NSFH model with $M=2363$ inducing points and $L=20$ components. A basic neuroanatomy diagram of mouse brain is provided for reference in Figure \ref{fig:vz-gof-val}b. Spatial importance scores indicate that spatial factors are most explanatory for variation at both the gene and observation level (Figure \ref{fig:vz-gof-val}c-d). Similar to the Slide-seqV2 data, most spatial factors mapped to specific brain regions (Figure \ref{fig:vz-npfh20-heatmap}a) such as the cerebral cortex (2), corpus callosum (4), and choroid plexus (10). Top genes for each spatial component again showed expression patterns overlapping with their associated factors (Figure \ref{fig:vz-npfh20-heatmap}b). While the majority of nonspatial factors were dispersed across the field-of-view (Figure \ref{fig:vz-npfh20-heatmap}c), a few of them did exhibit spatial localization to areas such as the hypothalamus (2) and hippocampus (4). This illustrates that the nonspatial factors are not antagonistic to spatial variation but should be thought of as spatially naive or agnostic. Given that Visium does not provide single-cell resolution, and this phenomenon was not observed in the other two datasets, we hypothesize that spatial patterns active in small numbers of observations may be more likely to be picked up as nonspatial factors under such conditions. Using the top genes for each component, we identified cell types and biological processes (Table \ref{table:vz-ctypes}). For example, spatial component 3 aligned to the basal ganglia, which is involved in the ``response to amphetamine'' biological process. Nonspatial component 10 had many genes associated with erythroid progenitor cells. The nonspatial patterns in this component suggest that this factor includes cell types in blood; however erythroid progenitor cells are not found in blood.. We hypothesize these are actually erythrocytes, which have been shown to retain parts of the erythroid transcriptome despite the loss of the nucleus \citep{doss_comprehensive_2015}. As in the Slide-seqV2 hippocampus data, neurons and glia were the most common cell types across all components.

\begin{table}[!htb]
\centering
\small
\caption{Nonnegative spatial factorization hybrid model (NSFH) identifies biologically distinct components in Visium mouse brain.}
\begin{adjustwidth}{-1in}{0in}
\begin{tabular}{|r|l|p{0.17\linewidth}|p{0.15\linewidth}|p{0.25\linewidth}|p{0.35\linewidth}|}
\hline
\textbf{dim} & \textbf{type} & \textbf{brain regions} & \textbf{cell types} & \textbf{genes} & \textbf{GO biological processes} \\ \hline
1 & spat & multiple &  & \textit{IGKC, COX6A2, TNNC1, CABP7, S100A9} & mitochondrial electron transport, NADH to ubiquinone, mitochondrial   respiratory chain complex I assembly \\ \hline
2 & spat & cerebral cortex & Interneurons & \textit{CCK, DKK3, STX1A, NRN1, RTN4R} & axonogenesis, positive regulation of behavioral fear response \\ \hline
3 & spat & basal ganglia & Neurons & \textit{GPR88, PDE10A, RGS9, PPP1R1B, PENK} & response to amphetamine, striatum development \\ \hline
4 & spat & fiber tracts/ corpus callosum & Oligodendrocytes & \textit{PLP1, MAL, MOBP, MAG, CLDN11} & myelination, central nervous system myelination \\ \hline
5 & spat & olfactory granule layer & Neurons & \textit{GNG4, GPSM1, CPNE4, SHISA8, MYO16} & embryonic limb morphogenesis, proximal/distal pattern formation \\ \hline
6 & spat & multiple & Interneurons & \textit{NPTXR, CCN3, SLC30A3, RASL10A, LMO3} & intracellular signal transduction, regulation of catalytic activity \\ \hline
7 & spat & outer olfactory bulb & Interneurons & \textit{S100A5, DOC2G, CDHR1, CALB2, FABP7} & mesoderm formation, cellular response to glucose stimulus \\ \hline
8 & spat & inner cerebral cortex & Neurons & \textit{HS3ST2, IGHM, CCN2, IGSF21, NR4A2} & isoprenoid biosynthetic process, cholesterol biosynthetic process \\ \hline
9 & spat & multiple & GABAergic neurons & \textit{GAD1, SLC32A1, HAP1, STXBP6, CPNE7} & neurotransmitter metabolic process, regulation of gamma-aminobutyric acid   secretion \\ \hline
10 & spat & choroid plexus of lateral ventricle & Choroid plexus cells & \textit{TTR, ECRG4, ENPP2, KL, 2900040C04RIK} & hormone transport, retinol metabolic process \\ \hline
1 & nsp &  & GABAergic neurons & \textit{PVALB, KCNAB3, CPLX1, VAMP1, SYT2} & positive regulation of potassium ion transmembrane transporter activity,   neuromuscular process \\ \hline
2 & nsp & hypothalamus & Neurons & \textit{BAIAP3, NNAT, HPCAL1, LYPD1, RESP18} & neurotrophin TRK receptor signaling pathway, muscle fiber development \\ \hline
3 & nsp &  & Neurons & \textit{NEFM, LGI2, DNER, PLCXD2, CARTPT} & regulation of synaptic vesicle fusion to presynaptic active zone membrane, neuronal action potential propagation \\ \hline
4 & nsp & hippocampus & Neurons & \textit{WIPF3, CPNE6, RGS14, ARPC5, CABP7} & Arp2/3 complex-mediated actin nucleation, calcineurin-NFAT signaling   cascade \\ \hline
5 & nsp &  & Neurons & \textit{HS3ST4, RAB26, CLSTN2, TLE4, SNCA} & vacuolar acidification, protein glycosylation \\ \hline
6 & nsp &  & Vascular fibroblasts & \textit{LARS2, GM42418, VTN, PTN, NPY} & establishment of epithelial cell polarity, plasma lipoprotein particle   organization \\ \hline
7 & nsp &  & Neurons & \textit{PLXND1, VSTM2L, CALB1, RGS7, MGAT5B} & heterophilic cell-cell adhesion via plasma membrane cell adhesion   molecules, short-term memory \\ \hline
8 & nsp &  & GABAergic neurons & \textit{SST, NPY, RESP18, NOS1, PDYN} & neuropeptide signaling pathway, regulation of the force of heart   contraction \\ \hline
9 & nsp &  & Astrocytes & \textit{TUBB2B, GM3764, NTRK2, MFGE8, PLPP3} & complement activation, negative regulation of growth \\ \hline
10 & nsp &  & Erythrocytes & \textit{HBA-A1, HBA-A2, HBB-BT, HBB-BS, ALAS2} & oxygen transport, cellular oxidant detoxification \\ \hline
\end{tabular}
\end{adjustwidth}
\label{table:vz-ctypes}
\end{table}

\clearpage

\section{Discussion}
We have presented nonnegative spatial factorization (NSF), a probabilistic approach to spatially-aware dimension reduction on observations of count data based on Gaussian processes (GP). We also showed how to combine spatial and nonspatial factors with the NSF hybrid model (NSFH). On simulated data, NSF, NSFH, and the nonspatial model probabilistic NMF (PNMF) all recovered an interpretable parts-based representation, whereas real-valued factorizations such as MEFISTO \citep{velten_identifying_2020} led to a less interpretable embedding. A key advantage of spatially-aware factorizations over unsupervised alternatives such as factor analysis (FA) and PNMF is generalizability; spatial factor models learn latent functions over the entire spatial domain rather than only at the observed locations. On a benchmarking task using three spatial transcriptomics (ST) datasets from three different technologies, NSF and NSFH had consistently lower out-of-sample prediction error than PNMF. We found that using a Matérn kernel reduced prediction error in our implementation of real-valued spatial factorization (RSF) when compared to MEFISTO, which uses a squared exponential kernel. We demonstrated how NSFH spatial and nonspatial components identify distinct regions in brain and liver tissue, cell types, and biological processes. Finally, we quantified the proportion of variation explained by spatial versus nonspatial components at both the gene and observation level using spatial importance scores.

A substantial limitation of the models studied here is the reliance on Euclidean distance as the metric for the GP kernel over the spatial domain. While this was appropriate for the particular datasets we explored, other ST datasets more closely resemble manifolds. An example is the embryonic tissue profiled by \cite{lohoff_highly_2020}.
Under such conditions, standard GP kernels are inappropriate \citep{dunson_diffusion_2020}. The recently proposed manifold GP \citep{borovitskiy_matern_2020} and graph GP \citep{borovitskiy_matern_2021} seem promising as alternatives. Either of these could be substituted for the standard GPs in our RSF, NSF, and NSFH models.

Historically, two major challenges for working with ST data have included integration with single-cell RNA-seq references \citep{lopez_joint_2019,verma_bayesian_2020} and deconvolving observations that incorporate multiple cells \citep{cable_robust_2021,lopez_multi-resolution_2021}. Addressing these will be an important future direction for research into nonnegative spatial factor models. However, we anticipate that ongoing improvements in ST protocols will increase the number of genes detected per location while improving the spatial resolution to single-cell or even subcellular levels \citep{eng_transcriptome-scale_2019,xia_spatial_2019} while maintaining a wide field-of-view.

All of the spatial models we considered were based on linear combinations of GPs with variational inference using inducing points \citep{leibfried_tutorial_2021,van_der_wilk_framework_2020}. While this technique has greatly improved GP scalability by enabling minibatching and nonconjugate likelihoods, the computational complexity still scales cubically with the number of inducing points. Promising future directions for GP inference include the harmonic kernel decomposition \citep{sun_scalable_2021}, nearest-neighbor GPs \citep{finley_efficient_2019}, and random Fourier features \citep{hensman_variational_2017,gundersen_latent_2020}. While linearity and nonnegativity are advantageous for interpretability \citep{svensson_interpretable_2020}, multivariate spatial factor models can also be formulated using nonlinear deep GPs \citep{wu_brain_2021}.

We have focused on the application of nonnegative spatial factor models to genomics data. However, both NSF and NSFH are relevant to other types of multivariate spatial or temporal data. Examples include forestry and environmental remote sensing \citep{taylor-rodriguez_spatial_2019}, wearable devices \citep{straczkiewicz_systematic_2021}, and neuroscience \citep{wu_gaussian_2017,foti_statistical_2019}.

\section{Methods}
\subsection{Nonnegative spatial factorization inference}
\subsubsection{Evidence lower bound objective (ELBO) function}
The posterior distribution of NSF cannot be computed in closed form, so we resort to approximate inference using a variational distribution. We assume a set of inducing point locations $z_m$ indexed by $m=1,\ldots,M$. If $M=N$ we set $z_m$ to be the spatial coordinates $X$. Otherwise, for $M<N$ we set $z_m$ to be the center points of a k-mean clustering (with $k=M$) applied to $X$. Let $u_{ml}=f_l(z_m)$ be the inducing points, i.e., the Gaussian process evaluation of the inducing locations. We are interested in inference of the posterior of the latent variables $u_{ml}$ and $f_{il}$. Temporarily assume loadings $w_{jl}$, likelihood shape and dispersion parameters, GP prior mean parameters $\beta_l$, and kernel hyperparameters $\theta_l$ are known. The posterior is given by
\begin{align*}
p(U,F\vert~Y;~X,Z) &= \frac{p(Y\vert~U,F)p(U,F;~X,Z)}{\int_{U,F}p(Y\vert~U,F)p(U,F;~X,Z)}\\
&= \frac{p(Y\vert~F)p(F\vert~U;~X,Z)p(U;~Z)}{\int_{U,F}p(Y\vert~F)p(F\vert~U;~X,Z)p(U;~Z)}.
\end{align*}
Note that the likelihood term depends on $U$ only through $F$, and we have decomposed the joint prior on $U,F$ into a marginal prior of $U$ and a conditional prior of $F\vert~ U$.
Following \cite{salimbeni_doubly_2017} and \cite{van_der_wilk_framework_2020}, the GP prior for inducing points is given by
\begin{align*}
p(U;~Z)&=\prod_{l=1}^L p(\vec{u}_l;~Z)\\
p(\vec{u}_l;~Z)&=\mathcal{N}\big(\mu_l(Z),~K_{uul}\big)\\
\big[K_{uul}\big]_{m,m'}&=k_l\big(\vec{z}_m,\vec{z}_{m'}).
\end{align*}
Next, we specify the GP prior for the function values at the observed locations by conditioning on the inducing points.
\begin{align*}
p(F\vert~U;~X,Z) &= \prod_{l=1}^L p(\vec{f}_l\vert~\vec{u}_l;~X,Z)\\
p(\vec{f}_l\vert~\vec{u}_l;~X,Z) &= \mathcal{N}\big(\vec{\mu}_{f\vert~ ul},~K_{f\vert~ ul}\big)\\
\vec{\mu}_{f\vert~ ul} &= \mu_l(X)+K_{ufl}'K_{uul}^{-1}\big(\vec{u}_l-\mu_l(Z)\big)\\
K_{f\vert~ ul} &= K_{ffl} - K_{ufl}'K_{uul}^{-1}K_{ufl}\\
\big[K_{ufl}\big]_{m,i} &= k_l(\vec{z}_m,\vec{x}_i)\\
\big[K_{ffl}\big]_{i,i'} &= k_l(\vec{x}_i,\vec{x}_{i'}).
\end{align*}
Note that $K_{uul}\in\mathbb{R}^{M\times M}$, $K_{ffl}\in\mathbb{R}^{N\times N}$, and $K_{ufl}\in\mathbb{R}^{M\times N}$.

We use the following approximation to the true posterior to facilitate variational inference:
\begin{align*}
q(U,F;~X,Z) &= p(F\vert~ U;~X,Z)q(U;~Z)\\
q(U;~Z) &= \prod_{l=1}^L q(\vec{u}_l;~Z)\\
q(\vec{u}_l;~Z)&=\mathcal{N}\big(\vec{\delta}_l,~\Omega_l\big).
\end{align*}
We will later need to draw samples of $F$ from this distribution. This is made easier by analytically marginalizing out $U$.
\begin{align*}
q(\vec{f}_l\vert~\vec{\delta}_l,\Omega_l;X,Z) &= \int_{\vec{u}_l}p(\vec{f}_l\vert~ \vec{u}_l;~X,Z)q(\vec{u}_l;~Z)\\
&= \mathcal{N}\big(\tilde{\mu}_l,~\tilde{\Sigma}_l),
\end{align*}
where the marginal mean vector $\tilde{\mu}_l\in\mathbb{R}^N$ and covariance matrix $\tilde{\Sigma}_l\in\mathbb{R}^{N\times N}$ are given by
\begin{align*}
\tilde{\mu}_l &= \mu_l(X) + K_{ufl}'K_{uul}^{-1}\big(\vec{\delta}_l-\mu_l(Z)\big)\\
\tilde{\Sigma}_l &= K_{ffl}-K_{ufl}'K_{uul}^{-1}\big(K_{uul}-\Omega_l\big)K_{uul}^{-1}K_{ufl}.
\end{align*}

Minimizing the KL divergence from the true posterior distribution to the approximating distribution is equivalent to maximizing the following evidence lower bound (ELBO)
\citep{hensman_gaussian_2013,salimbeni_doubly_2017,van_der_wilk_framework_2020}:
\begin{align*}
\mathcal{L} &= \E_{q(U,F)}\left[\log\frac{p(Y\vert~F)p(F\vert~U;~X,Z)p(U;~Z)}{q(U,F)}\right]\\
&= \E_{q(U,F)}\big[\log p(Y\vert~F)\big] + \E_{q(U,F)}\left[\log\frac{p(F\vert~U;~X,Z)p(U;~Z)}{p(F\vert~ U;~X,Z)q(U;~Z)}\right]\\
&= (\mathcal{L}_1) - \sum_{l=1}^L \E_{q(\vec{u}_l)}\left[\log\frac{q(\vec{u}_l;~Z)}{p(\vec{u}_l;~Z)}\right]\\
&= \mathcal{L}_1 - \sum_{l=1}^L \KL\big(q(\vec{u}_l)~\vert\vert~p(\vec{u}_l)\big).
\end{align*}
The KL divergence term from prior to approximate posterior has a closed-form expression since both are Gaussian (recall $M$ is the total number of inducing points):
\begin{align*}
\KL\big(q(\vec{u}_l)~\vert\vert~p(\vec{u}_l)\big) &= \frac{1}{2}\left[\log\frac{\vert K_{uul}\vert}{\vert \Omega_l \vert} - M + \tr\big\{K_{uul}^{-1}\Omega_l\big\} + \big(\vec{\delta}_l-\mu_l(Z)\big)'K_{uul}^{-1}\big(\vec{\delta}_l-\mu_l(Z)\big)\right].
\end{align*}
Let $\zeta(y\vert~\nu\lambda)$ be the log likelihood of an exponential family such as the Poisson or negative binomial distribution with mean $\nu\lambda$. In particular, for the Poisson distribution, $\zeta(y\vert~\nu\lambda)=y\log(\nu\lambda)-\nu\lambda-\log y!$. Let $F[i,:]=(f_{i1},\ldots,f_{iL})$. The expected log likelihood term in the ELBO is given by:
\begin{align*}
\mathcal{L}_1 &= \sum_{i=1}^N\sum_{j=1}^J \E_{q(U,F)}\left[\zeta(y_{ij}\vert~\nu_i\lambda_{ij})\right]\\
&= \sum_{i=1}^N\sum_{j=1}^J \E_{q(F)} \left[\zeta\left(y_{ij}\left\vert~ \sum_{l=1}^L w_{jl}e^{f_{il}}\right.\right)\right]\\
&= \sum_{i=1}^N\sum_{j=1}^J \E_{q(F[i,:])} \left[\zeta\left(y_{ij}\left\vert~ \sum_{l=1}^L w_{jl}e^{f_{il}}\right.\right)\right].
\end{align*}
The expectation in the above equation is intractable due to the nonlinear log likelihood function $\zeta(\cdot)$. However, we can simplify it in two ways. First, it only depends on $U$ through $F$, so the marginalized distribution $q(F)$ may be used instead of $q(U,F)$. Second, the log likelihood only depends on the marginal $f_{il}$ terms, as opposed to the multivariate $\vec{f}_l=(f_{1l},\ldots,f_{Nl})$ or the multivariate $F[i,:]$.
The approximate posterior distribution is therefore $q(F[i,:]) = \prod_{l=1}^L q(f_{il}) = \prod_{l=1}^L\mathcal{N}\left(\left[\tilde{\mu}_l\right]_i,~\left[\tilde{\Sigma}_l\right]_{i,i}\right)$, where
\begin{align*}
\alpha_l(\vec{x}_i) &= K_{uul}^{-1}\big[K_{ufl}\big]_{:,i}\\
\left[\tilde{\mu}_l\right]_i &= \mu_l(\vec{x}_i)+\alpha_l(\vec{x}_i)'\big(\vec{\delta}_l-\mu_l(Z)\big)\\
\left[\tilde{\Sigma}_l\right]_{i,i} &= k_l(\vec{x}_i,\vec{x}_i)-\alpha_l(\vec{x}_i)'\big(K_{uul}-\Omega_l\big)\alpha_l(\vec{x}_i).
\end{align*}
Despite these simplifications, the expectation still lacks a closed-form solution and is evaluated by approximation using Monte Carlo (MC) sampling \citep{salimbeni_doubly_2017}. The MC procedure draws $S$ samples $f^{(s)}_{il}\sim \mathcal{N}\left(\left[\tilde{\mu}_l\right]_i,~\left[\tilde{\Sigma}_l\right]_{i,i}\right)$ then evaluates
\[\E_{q(F[i,:])} \left[\zeta\left(y_{ij}\left\vert~ \sum_{l=1}^L w_{jl}e^{f_{il}}\right.\right)\right]\approx \frac{1}{S}\sum_{s=1}^S \left[\zeta\left(y_{ij}\left\vert~ \sum_{l=1}^L w_{jl}e^{f^{(s)}_{il}}\right.\right)\right]\].
In practice, we found $S=3$ to provide a reasonable balance between speed and numerical stability.

\subsubsection{Parameter estimation}
Using the ELBO as an objective function, we optimize all parameters using the Adam algorithm \citep{kingma_adam:_2014} with gradients computed by automatic differentiation in Tensorflow \citep{abadi_tensorflow_2016}. This includes the loadings weights $w_{jl}$, mean function intercepts $\beta_{0l}$ and slopes $\vec{\beta}_{1l}$, kernel length scale and amplitude parameters, variational location $\vec{\delta}_l$ and covariance $\Omega_l$ parameters, and any shape or dispersion parameters associated with the likelihood (e.g., for negative binomial and Gaussian distributions). 

To satisfy the nonnegativity constraint on $w_{jl}$, we use a projected gradient approach. After each optimization step, any values that are negative are truncated to zero.
All other parameter constraints are accommodated by monotone transformations. For example, the variational covariance matrices $\Omega_l$ must all be positive definite, so we store and use the lower triangular Cholesky decomposition factors instead of the full covariance matrices themselves.

\subsubsection{Real-valued spatial factorization inference}

The inference procedure for RSF is identical to NSF except we do not exponentiate the sampled $f_{il}^{(s)}$ terms prior to combining with the loadings $w_{jl}$. Because the loadings are no longer constrained to be greater than or equal to zero, the truncation step is omitted during optimization. To facilitate comparisons with MEFISTO, we focused on a Gaussian likelihood and only applied RSF to normalized data with features centered to have zero mean.

\subsection{Nonspatial count factorization inference}

To fit PNMF and FA models, we adopt a mean field variational approximation \citep{blei_variational_2016} to the posterior distribution of the latent factors:
\[q(f_{il}) = \mathcal{N}(\delta_{il},~\omega_{il})\].
Focusing on PNMF, the ELBO is of the form
\begin{align*}
\mathcal{L} &= \sum_{i=1}^N\sum_{j=1}^J \E_{q(F[i,:])} \left[\zeta\left(y_{ij}\left\vert~ \sum_{l=1}^L w_{jl}e^{f_{il}}\right.\right)\right] - \sum_{i=1}^N\sum_{l=1}^L \KL\big(q(f_{il})~\vert\vert~p(f_{il})\big).
\end{align*}
The expectation in the first term is approximated by MC sampling just as in NSF. The second term involves two univariate Gaussians and has the closed form
\begin{align*}
\KL\big(q(f_{il})~\vert\vert~p(f_{il})\big) &= \frac{1}{2}\left[\log\frac{s^2_l}{\omega_{il}} - 1 + \frac{\omega_{il}}{s^2_l} +\frac{(\delta_{il}-m_l)^2}{s^2_l}\right].
\end{align*}
FA has an identical setup to NSF except without exponentiating the sampled $f_{il}^{(s)}\sim q(f_{il})$. Optimization of parameters is the same as in NSF, including the truncation of $w_{jl}$ terms in PNMF.

\subsection{Nonnegative spatial factorization hybrid (NSFH) model}

Recall $\nu_i\lambda_{ij}$ is the mean of the outcome $y_{ij}$, which we assume is distributed as some exponential family likelihood, such as Gaussian, negative binomial, or Poisson.
The NSFH model is specified as the combination of $T$ spatial factors with $L-T$ nonspatial factors
\[\lambda_{ij} = \sum_{l=1}^T w_{jl}e^{f_{il}} + \sum_{l=T+1}^L v_{jl}e^{h_{il}}.\]
To estimate the $f_{il}$ terms, we use the same GP prior and variational inducing point approximate posterior as in NSF. To estimate the $h_{il}$ terms, we use the same univariate Gaussian prior and mean field variational approximate posterior as in PNMF. The ELBO objective function is similar to NSF and PNMF:
\begin{align*}
\mathcal{L} &= \sum_{i=1}^N\sum_{j=1}^J \E_{q(F[i,:],H[i,:])} \left[\zeta\left(y_{ij}\left\vert~ \sum_{l=1}^L w_{jl}e^{f_{il}} + \sum_{l=T+1}^L v_{jl}e^{h_{il}}\right.\right)\right]\ldots\\
&\ldots -\sum_{l=1}^L \KL\big(q(\vec{u}_l)~\vert\vert~p(\vec{u}_l)\big) - \sum_{i=1}^N\sum_{l=1}^L \KL\big(q(h_{il})~\vert\vert~p(h_{il})\big).
\end{align*}
Due to the mean-field formulation, the variational distributions factorize over components:
\[q(F[i,:],H[i,:])=\prod_{l=1}^T q(f_{il}) \prod_{l=T+1}^L q(h_{il}).\]
Thus, we approximated the expectation by independent MC samples of $f_{il}^{(s)}\sim q(f_{il})$ and $h_{il}^{(s)}\sim q(h_{il})$. The remaining two KL divergence terms are identical to those in NSF and PNMF and have the same closed form. We optimize all parameters using the same techniques described for NSF and PNMF.

\subsection{Postprocessing nonnegative factorizations}
Consider a generic nonnegative factorization $\Lambda=FW'$ or equivalently $\lambda_{ij}=\sum_l f_{il}w_{jl}$. We assume that the log-likelihood of data $Y$ depends on the $N\times L$ factors matrix $F$ and $J\times L$ loadings matrix $W$ only through $\Lambda$. The number of observations is $N$, number of components is $L$, and number of features is $J$. For notational simplicity, here we use $f_{il}$ to denote a nonnegative entry of a factor matrix rather than $e^{f_{il}}$ used in other sections. In the case that the model is probabilistic, we assume $f_{il}$ represents a posterior mean, posterior geometric mean, or other point estimate. We project $F,W$ onto the simplex while leaving the likelihood invariant.
\begin{align*}
\bar{f} &= \sum_{i=1}^N F_{[i,:]}\in \mathbb{R}^L\\
F&\gets F*\diag(\bar{f})^{-1}\\
W&\gets W*\diag(\bar{f})\\
\bar{w} &= \sum_{l=1}^L W_{[:,l]}\in \mathbb{R}^J\\
W&\gets \diag(\bar{w})^{-1}*W.
\end{align*}
Note that after this transformation $\Lambda = FW'\diag(\bar{w})$. We now have that the columns of $F$ all sum to one and the rows of $W$ all sum to one (i.e., they lie on the simplex). In the spatial transcriptomics context, the features are genes. A particular row of $W$ represents a single gene's soft clustering assignment to each of the $L$ components. If $w_{jl}=1$ this meant all of that gene's expression could be predicted using only component $l$, whereas if $w_{jl}=0$ this meant that component $l$ was irrelevant to gene $j$. For a given component $l$, we identified the top associated genes by sorting the $w_{jl}$ values in decreasing order.

We refer to the above procedure as ``SPDE-style'' postprocessing due to its similarity to spatialDE \citep{svensson_spatialde_2018}. An alternative postprocessing scheme is ``LDA-style'' \citep{blei_latent_2003,carbonetto_non-negative_2021} where the roles of $F$ and $W$ are switched. This results in a loadings matrix whose columns sum to one (``topics'') and a factors matrix whose rows sum to one. LDA-style postprocessing provides a soft clustering of observations instead of features. We used SPDE-style postprocessing throughout this work with the sole exception of computing spatial importance scores for observations, described below.

\subsubsection{NSFH spatial importance scores}
Let $F\in\mathbb{R}^{N\times T}_+$ represent the spatial factors matrix (rather than $e^F$), with corresponding loadings $W\in\mathbb{R}^{J\times T}_+$. Similarly let $H\in\mathbb{R}^{N\times (L-T)}_+$ represent the nonspatial factors (rather than $e^H$) with corresponding loadings $V\in\mathbb{R}^{J\times (L-T)}_+$. Let $A=[F,H]\in\mathbb{R}^{N\times L}_+$ and $B=[W,V]\in\mathbb{R}^{J\times L}_+$.

To obtain spatial importance scores for features (genes), we applied SPDE-style postprocessing to $A,B$. The score $\gamma_j$ for feature $j$ is given by the sum of the loadings weights across all the spatial components.
\begin{align*}
W\gets B_{[:,1:T]}\\
\gamma_j = \sum_{l=1}^T w_{jl}.
\end{align*}
Due to the initial postprocessing, $0\leq \gamma_j\leq 1$ for all $j$. If $\gamma_j=0$ then all the variation in feature $j$ was explained by the nonspatial factors. If $\gamma_j=1$ then all the variation was explained by the spatial factors.

To obtain spatial scores for observations, we applied LDA-style postprocessing to $A,B$. The score $\rho_i$ for observation $i$ is given by the sum of the factor values across all the spatial components.
\begin{align*}
F\gets A_{[:,1:T]}\\
\rho_i = \sum_{l=1}^T f_{il}.
\end{align*}
As before, $0\leq \rho_i\leq 1$ for all $i$. If $\rho_i=0$ then all the variation in observation $i$ was explained by the nonspatial factors. If $\rho_i=1$ then all the variation was explained by the spatial factors.

\subsection{Initialization}
Real-valued models were initialized with singular value decomposition. Nonnegative models were initialized with the scikit-learn implementation of NMF \citep{pedregosa_scikit-learn_2011}. For NSFH, we sorted the initial NMF factors and loadings in decreasing order of spatial autocorrelation using Moran's I statistic \citep{moran_notes_1950} as implemented in squidpy \citep{palla_squidpy_2021}. The first $T$ factors were assigned to the spatial component and the remaining $L-T$ factors to the nonspatial component.

\subsection{Simulations}
In the ggblocks simulation, each latent factor representing a canonical spatial pattern consisted of a $30\times 30$ grid of locations ($900$ total spatial locations). The number of features (``genes'') was set to $500$. Each gene was randomly assigned to one of the four patterns with uniform probabilities. The $900\times 500$ mean matrix was defined as the sum of two nonnegative matrices: one spatial ($M_1$) and one nonspatial ($M_2$). Entries of $M_1$ were set to $11$ in the active region (where a shape is visible) and $0.1$ elsewhere. We then generated a set of three nonspatial factors each of length $900$ by drawing from a Bernoulli distribution with probability $0.2$. Each gene was randomly assigned to one of the three nonspatial factors with uniform probabilities. The entries of $M_2$ were then set to $9$ in active cells and $0.1$ elsewhere. The counts were then drawn from a negative binomial distribution with mean $M_1+M_2$ and shape parameter $10$ to promote overdispersion. For MEFISTO, RSF, and FA the count data were normalized to have the same total count at each spatial location, then log transformed with a pseudocount of one. Features were centered before applying each dimension reduction method. For PNMF, NSF, and NSFH the raw counts were used as input. The unsupervised methods (PNMF and FA) used only the $900\times 500$ count matrix, while the supervised methods (NSF, NSFH, RSF, and MEFISTO) also used the $900\times 2$ matrix of spatial coordinates. Since this was a smaller dataset, all spatial coordinates were used as inducing point locations to maximize accuracy. All models were fit with $L=4$ components, except NSFH which was fit with $L=7$ total components of which $T=4$ were spatial and the rest nonspatial.

For the quilt simulation, we followed the same procedure as above, but the spatial patterns were $36\times 36$, leading to $1296$ total observations, each at a unique spatial location.

\subsection{Data acquisition and preprocessing}
For all datasets, after quality control filtering of observations, we selected the top $2000$ informative genes using Poisson deviance as a criterion \citep{townes_feature_2019,street_scry_2021}. Raw counts were used as input to nonnegative models (NSF, PNMF, NSFH) with size factors computed by the default Scanpy method as described below \citep{wolf_scanpy_2017}. For real-valued models with Gaussian likelihoods (RSF, FA, MEFISTO), we followed the default Scanpy normalization for consistency with MEFISTO. The raw counts were normalized such that the total count per observation equaled the median of the total counts in the original data. The normalized counts were then log transformed with a pseudocount of one, and the features were centered to have mean zero. This scaled, log-normalized version of the data was then used for model fitting.
\subsubsection{Visium mouse brain}
The dataset ``Mouse Brain Serial Section 1 (Sagittal-Anterior)`` was downloaded from \url{https://support.10xgenomics.com/spatial-gene-expression/datasets}. To facilitate comparisons, preprocessing followed the MEFISTO tutorial (\url{https://nbviewer.jupyter.org/github/bioFAM/MEFISTO_tutorials/blob/master/MEFISTO_ST.ipynb}) \citep{velten_identifying_2020}. Observations (spots) with total counts less than $100$ or mitochondrial counts greater than $20\%$ were excluded.
\subsubsection{Slide-seqV2 mouse hippocampus}
This dataset was originally produced by \cite{stickels_highly_2021}. We obtained it through the SeuratData R package \citep{satija_seuratdata_2019} and converted it to a Scanpy H5AD file \citep{wolf_scanpy_2017} using SeuratDisk \citep{hoffman_seuratdisk_2021}. Observations (spots) with total counts less than $100$ or mitochondrial counts greater than $20\%$ were excluded.
\subsubsection{XYZeq mouse liver}
This dataset was originally produced by \cite{lee_xyzeq_2021}. We obtained it from the Gene Expression Omnibus \citep{edgar_gene_2002}, accession number GSE164430. We focused on sample liver\_slice\_L20C1, which was featured in the original publication, and downloaded it as an H5AD file. The spatial coordinates were provided by the original authors. We did not exclude any observations (cells), since all had total counts greater than $100$ and mitochondrial counts less than $20\%$.

\subsection{Cell types and GO terms}
For each dataset, we fit a NSFH model and applied SPDE-style postprocessing such that the loadings matrices had rows (representing genes) summing to one across all components. We then examined each column of the loadings matrix (representing a component) and identified the five genes with largest weights. We then manually searched for cell types on scfind (\url{https://scfind.sanger.ac.uk/}) \citep{lee_fast_2021}. If no results were found, we next searched the Panglao database (\url{https://panglaodb.se}) \citep{franzen_panglaodb_2019}. We identified brain regions in the Slide-seqV2 hippocampus and Visium brain datasets by referring to the interactive Allen Brain Atlas (\url{https://atlas.brain-map.org}) \citep{wang_allen_2020}. GO annotations for all genes were downloaded from the BioMart ENSEMBL database (release 104, May 2021) using the biomaRt package in Bioconductor (version 3.13). Enriched terms were identified using the topGO Bioconductor package with the default algorithm ``weight01'' and statistic ``fisher'', considering the top $100$ genes for each component against a background of all other genes. 

\subsection{Software versions and hardware}
We implemented all models using Python 3.8.10, tensorflow 2.5.0, tensorflow probability 0.13.0. Other Python packages used include scanpy 1.8.0, squidpy 1.1.0, scikit-learn 0.24.2, pandas 1.2.5, numpy 1.19.5, scipy 1.7.0. We used the MEFISTO implementation from the mofapy2 Python package, installed from the GitHub development branch at commit 8f6ffcb5b18d22b3f44ff2a06bcb92f2806afed0.
Graphics were generated using either matplotlib 3.4.2 in Python or ggplot2 3.3.5 \citep{wickham_ggplot2_2016} in R (version 4.1.0). The R packages Seurat 0.4.3 \citep{hao_integrated_2021}, SeuratData 0.2.1, and SeuratDisk 0.0.0.9019 were used for some initial data manipulations.
Computationally-intensive model fitting was done on Princeton's Della cluster. Each model was assigned 12 CPU cores. We provided the following total memory per dataset: 180 Gb for Slide-seq V2, 72 Gb for Visium, and 48 Gb for XYZeq.

Code for reproducing the analyses of this manuscript is available from \url{https://github.com/willtownes/nsf-paper}. 

\acks{
The authors thank Britta Velten and the MEFISTO team for assistance in model fitting and responsiveness in bug fixes. George Hartoularos and Dylan Cable gave helpful advice on data acquisition and processing. Adelaide Minerva contributed expertise in neuroanatomy. Tina Townes helped with graphics. Anqi Wu and Stephen Keeley, along with members of the ``BEEHIVE'' including Archit Verma, Didong Li, Andy Jones and Siena Dumas Ang provided valuable insight through informal discussions. F.W.T. and B.E.E. were funded by a grant from the Helmsley Trust, a grant from the NIH Human Tumor Atlas Research Program, NIH NHLBI R01 HL133218, and NSF CAREER AWD1005627.

The work reported on in this paper relied on Princeton Research Computing resources at Princeton University, which is a consortium led by the Princeton Institute for Computational Science and Engineering (PICSciE) and the Office of Information Technology's Research Computing.
}

\bibliography{references}
\clearpage

\section*{Supplemental Figures}%
\renewcommand{\thefigure}{S\arabic{figure}}
\setcounter{figure}{0}

\begin{figure}[!htb]
\centering
\subfloat[ground truth]{
  \includegraphics[width=.49\linewidth]{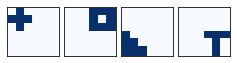}
}
\subfloat[simulated count data]{
  \includegraphics[width=.49\linewidth]{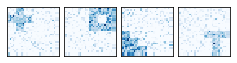}
}
\hspace{0em}
\subfloat[factor analysis]{
  \includegraphics[width=.49\linewidth]{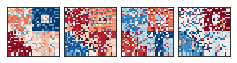}
}
\subfloat[probabilistic nonnegative matrix factorization]{
  \includegraphics[width=.49\linewidth]{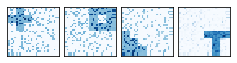}
}
\hspace{0em}
\subfloat[MEFISTO]{
  \includegraphics[width=.49\linewidth]{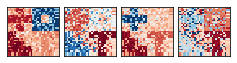}
}
\subfloat[real-valued spatial factorization]{
  \includegraphics[width=.49\linewidth]{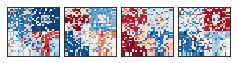}
}
\hspace{0em}
\subfloat[nonnegative spatial factorization]{
  \includegraphics[width=.49\linewidth]{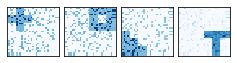}
}
\subfloat[nonnegative spatial factorization hybrid model]{
  \includegraphics[width=.49\linewidth]{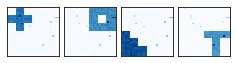}
}
\caption{\small Nonnegative factorizations recover a parts-based representation in ``ggblocks'' simulated multivariate spatial count data. (a) Each of $500$ features was randomly assigned to one of four nonnegative spatial factors. (b) Negative binomial count data used for model fitting. (c) Real-valued factors learned from unsupervised (nonspatial) dimension reduction. (d) as (c) but using nonnegative components. (e) Real-valued, spatially-aware factors with squared exponential kernel. (f) as (e) but with Matérn kernel. (g) Nonnegative, spatially-aware factors. (h) as (g) but with additional three nonspatial factors.}
\label{fig:ggblocks}
\end{figure}

\begin{figure}[!htb]
\centering
\subfloat[Sparsity]{
  \includegraphics[width=.49\linewidth]{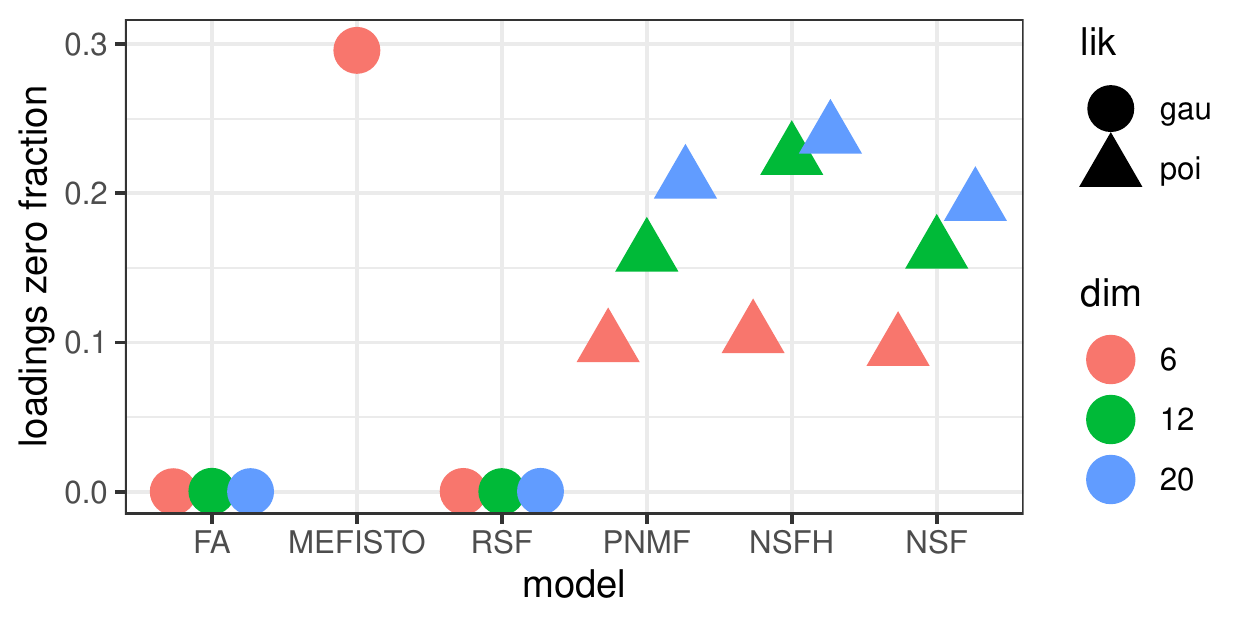}
}
\subfloat[Time to converge]{
  \includegraphics[width=.49\linewidth]{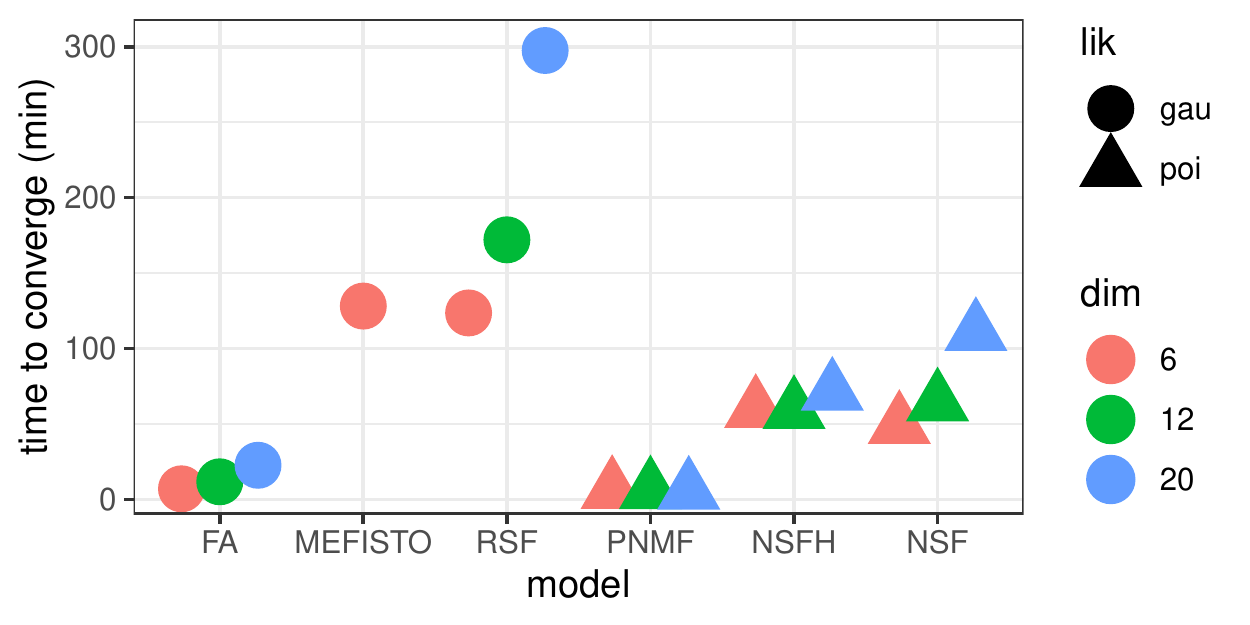}
}
\hspace{0em}
\subfloat[Validation deviance by likelihood]{
  \includegraphics[width=0.49\linewidth]{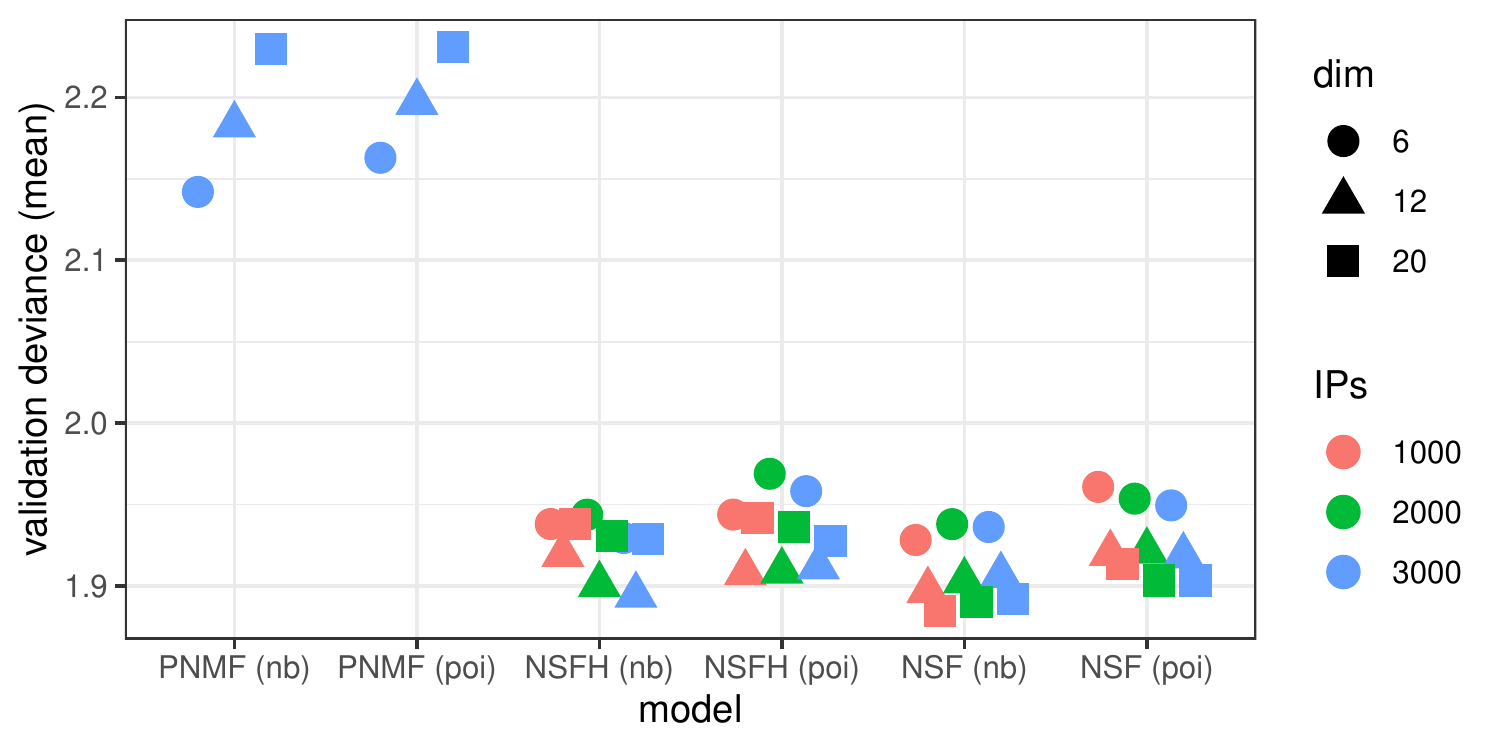}
}
\subfloat[Time to converge by likelihood]{
  \includegraphics[width=.49\linewidth]{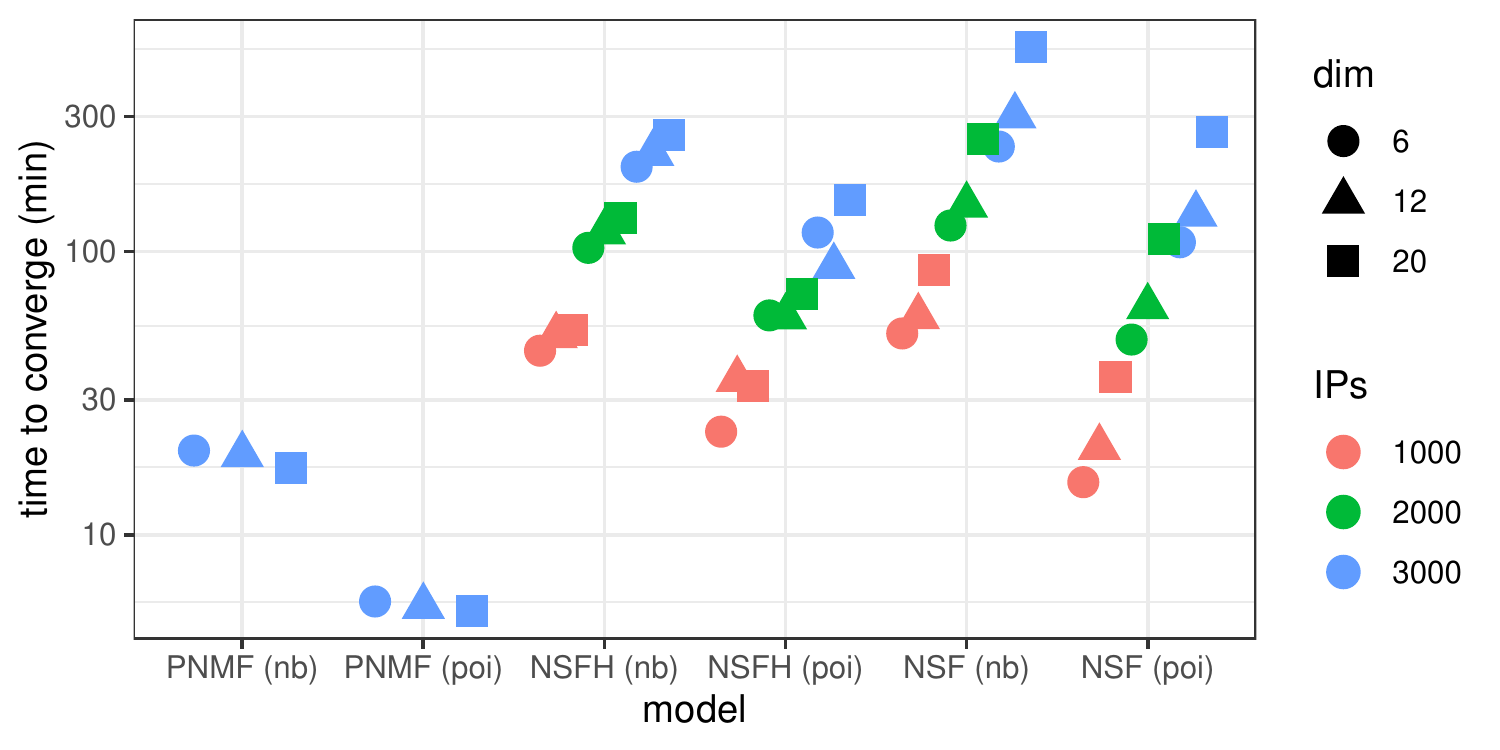}
}
\caption{\small Benchmarking spatial and nonspatial factor models on Slide-seqV2 mouse hippocampus gene expression data. FA: factor analysis, RSF: real-valued spatial factorization, PNMF: probabilistic nonnegative matrix factorization, NSF: nonnegative spatial factorization, NSFH: NSF hybrid model, lik: likelihood, gau: Gaussian, poi: Poisson, nb: negative binomial. (a) Sparsity of loadings matrix increases with larger numbers of components (dim) in nonnegative models PNMF, NSFH, and NSF. (b) Nonnegative spatial models NSF and NSFH converge faster than MEFISTO but not as fast as nonspatial models FA and PNMF. (c) Negative binomial and Poisson likelihoods provide similar generalization accuracy (lower deviance) in nonnegative models. (d) Negative binomial likelihood is more computationally expensive than Poisson likelihood in nonnegative models.}
\label{fig:sshippo-sparsity-timing}
\end{figure}

\end{document}